\documentclass{article}
\usepackage{graphicx} 
\usepackage{amsmath,amssymb,mathtools,stmaryrd} 
\usepackage[a4paper,bindingoffset=10mm,left=10mm,right=10mm,top=10mm,bottom=20mm,footskip=10mm]{geometry}
\usepackage{authblk}

\def\onedot{$\mathsurround0pt\ldotp$}
\def\cddot{
	\mathbin{\vcenter{\baselineskip.67ex
			\hbox{\onedot}\hbox{\onedot
			}
		}
	}
}
\newcommand{\norm}[1]{ \left\| #1 \right\| }
\newcommand{\I}{\mathbb{I}}
\newcommand{\R}{\mathbb{R}}
\usepackage{tikz}
\usetikzlibrary{calc}
\usetikzlibrary{arrows}
\usetikzlibrary{arrows.meta}
\usetikzlibrary{shapes}
\usetikzlibrary{angles}
\usetikzlibrary{positioning}
\usetikzlibrary{decorations}
\usetikzlibrary{3d}
\usetikzlibrary{quotes}
\usepackage{pgfplots}
\pgfplotsset{compat=1.11}
\usepgfplotslibrary{fillbetween}
\usetikzlibrary{intersections}
\usetikzlibrary{decorations.text}
\pgfdeclarelayer{bg}
\pgfsetlayers{bg,main}
\usepackage{tikz-3dplot} 
\usepackage{xifthen}

\usetikzlibrary{arrows.meta,
	calc, chains,
	quotes,
	positioning,
	shapes.geometric}

\tikzstyle{reverseclip}=[insert path={(current page.north east) --
	(current page.south east) --
	(current page.south west) --
	(current page.north west) --
	(current page.north east)}
]

\pgfplotsset{compat=newest} 
\usepgfplotslibrary{units} 
\usepackage{graphicx} 
\usepackage{color} 
\usepackage{caption}
\usepackage{subcaption}

\definecolor{ao}{rgb}{0.0, 0.5, 0.0}
\definecolor{applegreen}{rgb}{0.55, 0.71, 0.0}

\title{
On computing the jump condition of the dissipation rate in the two-equation turbulence models for two-phase flow and application to air-water waves 
}
\author[1]{Omar ELSAYED\thanks{First author}}
\author[1]{Benjamin Bouscasse \thanks{Corresponding author: \texttt{benjamin.bouscasse@ec-nantes.fr}}}
\author[2]{Maïté Gouin}
\author[1]{David Le Touzé}
\affil[1]{École Centrale Nantes, CNRS, LHÉEA, UMR 6598, 44000 Nantes, France}
\affil[2]{SIREHNA, Technocampus Ocean, 44340 Bouguenais, France}
\date{\today}
\begin{document}
\maketitle

\begin{abstract}
Traditional turbulence models are derived for single-phase flow. Extension of the family of two-equation turbulence models for two-phase flow is obtained via scaling the transport equations by the density. In the special case of two-phase flow with a sharp interface, jump conditions exist. Two types of jump conditions are found: (1) jump in the partial differential equation (PDE) physical quantities such as density and viscosity and (2) jump in the turbulence frequency. We first derive and clarify the jump in the equations. The jump in the turbulence frequency is proportional to the kinematic viscosity ratio, which is approximately $10$ in the case of air-water. Then a new field, the inverse turbulence area, is considered to model the turbulence effects instead of the turbulence frequency. For the system of air and water, the effect of the jump of the kinematic viscosity is always greater than the effect arising from the jump of velocity gradient. This approximation leads to the assumption of a continuous inverse turbulence area scale. Validation versus experimental measurements from the literature is then presented to demonstrate the improvement of the model. In particular, the wave breaking phenomenon is simulated in two conditions: spilling and plunging wave breakers. The proposed model shows its ability to predict the turbulence in the surf zone accurately. Finally, it explains the low values of the time-averaged turbulent kinetic energy in the surf zone which is caused by the increase of the turbulence frequency in the air. 
\end{abstract}
\section{Introduction}
Turbulence modeling in two-phase flow is vital to a wide spectrum of applications, e.g., coastal structures, offshore energy generation, wave added resistance and sea-keeping of ships. For example, in the case of a sailing naval structure, one needs to compute the water and air drag force and the wave added resistance. Wave motion is driven by the dynamics close to the interface; the magnitude of the velocity of the water or the air is negligible far away from the interface and the turbulence as well. However, Reynolds Averaged Navier Stokes (RANS) models are generally derived in a single-phase flow context. Thus, the properties of the interface poses a further complexity in the model.
\newline
At a fluid-fluid interface, a discontinuity of the physical quantities (density and viscosity) exists. This sharp jump leads to a hydrodynamic boundary condition i.e. a balance of normal and tangential stress at the interface. The shear stress at the interface depends on the dynamic viscosity and the velocity gradient. In order to balance the tangential stress, the projection of the velocity gradient in the tangent space of the interface exhibits a jump condition as well.
\newline
By fixing the mixing length at the interface to a characteristic value depending on the wave properties (wave crest, height, etc.), the turbulent time-scale undergoes a jump between the air and the water with one order of magnitude. Then, the turbulent wakes in the air are dissipated approximately in $0.1 $ the time needed to dissipate the wakes in the water. Moreover, from a mathematical point of view, both the convection and diffusion terms undergo a jump condition which describes the transport mechanism of the turbulence scales across the interface. It has been found in many researches that RANS models over-predict the turbulence levels especially in the water, see, e.g., \cite{lin_liu_1998,brown_evaluation_turb_surf_zone,mayer,hsu}.These works did not take into account the effect of that jump conditions.
\newline
Single equation models such as Prandtl mixing length model, \cite{prandtl}, Spalart-Allmaras, \cite{spalart}, etc. are not applied for the description of two-phase flow because of their incompleteness, see \cite{tennekes,pope}. These models rely on a predefined value of the turbulence length scale. The simplest complete description of turbulence requires at least two transport equations to describe the generation and dissipation of eddies. The fundamental idea behind the family of two-equation models is to describe both the mean turbulence kinetic energy (TKE) $k$ and the turbulence dissipation $\epsilon$ or frequency $\omega$ via two coupled reaction transport equations. The turbulence closure problem is to find loss and gain functions which model the generation and the destruction of turbulent eddies.
\newline
For the case of the $k-\epsilon$, \cite{launder,yakhot,rng} or $k-\omega$, \cite{menter_1, menter_2, wilcox}, models, the jump conditions are not considered. Devolder (\cite{devolder}) extended and tested the single phase models via multiplying the equations by the density. Furthermore, a buoyancy damping term was added in the TKE equation in order to damp and reduce the TKE around the interface (\cite{vanmale}). This modification leads to approximately laminar flow condition around the fluid-fluid interface and turbulent flow far away from it. Larsen and Fuhrman (\cite{larsen_fuhrman_2018}) showed that all the variants of both $k-\epsilon$ and $k-\omega$ are unconditionally unstable in the regions of potential flow. They devised a limiter function of the turbulence frequency in order to stabilize the model. Inspecting Figures $3$ and $4$  in \cite{larsen_fuhrman_2018} and figures $3$ and $6$ in \cite{devolder}, it is found that the turbulence kinetic energy follows a parabolic pattern. This pattern contradicts with the experimental measurements in \cite{ting}. In the experimental measurements, the TKE is linear rather than a smooth parabola. Furthermore, Figure 12 (except Sub-figures e and h) in \cite{larsen_fuhrman_2018} does not predict a steep decrease of the TKE in air as in the experimental measurements of \cite{ting}. They both used the open-source wave generation toolbox developed in \cite{waves2foam}.
\newline
In the present work we use the open-source package OpenFOAM to solve RANS equations \cite{of_paper,jasakPhD} in a multiphase flow context. The base solver is semi-implicit and it provides second order spatial discretization schemes. The in-house library foamStar, \cite{foamStar1}, is used to model the wave generation and boundary conditions. The volume of fluid (VoF) method, \cite{HIRT1981201}, is used for modeling the two phases where an algebraic advection scheme is used (variant of the face compression scheme in \cite{zalesakFCT}). The scheme utilizes a compression term to avoid smearing of the interface.
\newline
This paper is organized as follows: Section \ref{section:derivations} gives the integral forms of the turbulence model and the necessary analysis of the jump conditions across the interface. It also formulates a new two-equation turbulence model for two-phase flow with a sharp interface. Section \ref{sec:numerics} describes the main lines of the numerical discretization used in OpenFOAM. 
Section  \ref{sec:results} reports the simulation results for spilling and plunging breakers using the proposed model. The results are compared to experimental measurements in \cite{ting}. Section \ref{sec:conclusions} draws conclusions regarding the presented model and the associated CFD results.

\section{Derivation of the jump conditions}
\label{section:derivations}
\subsection{Problem of the fluid-fluid interface}
\begin{figure}
	\scalebox{1}{
		\begin{tikzpicture}[line join=blevel,z=0,scale=5]
	
	\coordinate (A1) at (0,0);
	\coordinate (A2) at (1,0);
	\coordinate (A3) at (1,1);
	\coordinate (A4) at (0,1);
	
	\coordinate (C1) at (0.5,0.5);
	\coordinate (C2) at (0.05,0.7);
	\coordinate (C3) at (0.2,0.5);
	\coordinate (C4) at (0,0.5);
	
	\draw [line width=2 pt,ao,fill=applegreen] (A1)--(A2)--(A3)--(A4)--cycle;
	\draw [red,fill=red]  (C1) circle (0.2);
	
	\node at (C1) [right,white] (TextNode) {$\Omega_2$};
	\node at (C2) [right,black] (TextNode) {$\Omega_1$};
	\node at (C3) [right,black] (TextNode) {$\Gamma$};
	\node at (C4) [left,black] (TextNode)  {$\partial \Omega $};
\end{tikzpicture}
	}
	\caption{Schematic of a two immiscible-fluid problem.}
	\label{vof_models}
\end{figure}
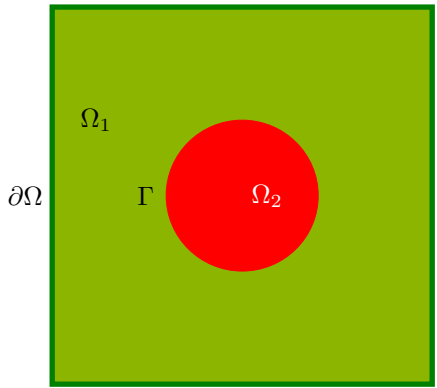

Consider a subset of the Euclidan space ($\Omega \in \R^m$, $m={2,3}$), which is divided by a sharp interface $\Gamma$ into two non-overlapping regions namely $\Omega_1$ and $\Omega_2$ as shown in Figure \ref{vof_models}. The interface induces a geometric basis $\left\{\hat{t},\hat{b},\hat{n}\right\}$ for the tangent, bi-normal and normal unit vectors, respectively. The unit normal vector is pointing outward the first phase to the second phase. The tangent and bi-normal unit vectors are arbitrary orthogonal unit vectors in the tangent space.
Each region (phase) is an incompressible fluid with density $\rho_{1,2}$ and dynamic viscosity $\mu_{1,2}$. The physical properties are constant in each region (phase). The jumps in the physical properties are
\begin{equation}
	\llbracket \rho  \rrbracket = \rho_2- \rho_1, \quad
	\llbracket \mu   \rrbracket = \mu_2 - \mu_1,\quad
	\llbracket \nu   \rrbracket = \nu_2 - \nu_1
	= \frac{\mu_2}{\rho_2}-\frac{\mu_1}{\rho_1}.
\end{equation}

\subsection{$K-\omega$ turbulence models}
Let the velocity vector field $u$ be decomposed into mean (time average velocity over a time interval much greater than the turbulence time scale) $\overline{u}$ and perturbation $u^\prime$, and similarly for the pressure field $p$ into mean $\overline{p}$ and perturbation $p^\prime$ as follows
\begin{equation}
	u = \overline{u} + u^\prime, \quad p = \overline{p} + p^\prime.
\end{equation}
The instantaneous turbulent kinetic energy reads
\begin{equation}
	k=\frac{1}{2} {u^\prime \cdot u^\prime}.
\end{equation}
The shear-strain rate is
\begin{equation}
	S=\frac{1}{2}\left( \nabla u + \nabla u^T \right).
\end{equation}
The turbulent (Reynolds) stress is defined as follows
\begin{equation}
	\tau_{t}=2 \nu_t S-\frac{2}{3}k \I,
\end{equation}
where $\nu_t$ is the turbulent eddy viscosity and $\I$ is the identity matrix. 
The governing equation for the mean turbulent kinetic energy reads
\begin{equation}\label{kEqn}
	\frac{\partial k}{\partial t}
	+\nabla \cdot \left(u k\right)
	=
	\tau_{t} \cddot \nabla u
	- \beta^\ast k \omega
	-\nabla \cdot \left(\left(\nu + \sigma^\ast \nu_t \right) \nabla k \right)
	,
\end{equation}
where $\cddot$ is the operator for double inner product, $\nu$ is the kinematic viscosity, $\tau_t$ is the turbulent shear stress and $\beta^\ast$ and $\sigma^\ast$ are model parameter and $\omega$ is the specific dissipation rate per unit mass. The specific dissipation rate per unit mass is governed by 
\begin{equation}\label{omegaEqn}
	\frac{\partial \omega}{\partial t}
	+\nabla \cdot \left(u \omega\right)
	=
	\frac{\gamma \omega}{k}
	- \beta \omega^2
	-\nabla \cdot \left( \left(\nu + \sigma \nu_t \right) \nabla \omega \right)
	+F  \frac{\sigma_\omega}{\omega} \nabla k \cdot \nabla \omega 
	,
\end{equation}
where
\begin{equation}\label{eddyViscosity}
	\nu_t
	=
	\frac{k}{\omega},
	\quad
	\mu_t = \rho \nu_t = \rho \frac{k}{\omega},
\end{equation}
where $\beta$ and $\sigma$ are model parameters. The previous set of Equations \ref{kEqn} and \ref{omegaEqn} are usually re-written in scaled form i.e. multiplied by the density ($\rho$) which is constant in the region occupied by an incompressible fluid as follows
\begin{equation}\label{rhoKEqn}
	\begin{aligned}
		&\frac{\partial \rho k}{\partial t}
		+\nabla \cdot \left(\rho u k\right)
		=
		\rho \tau_{t} \cddot \nabla u
		- \beta^\ast \rho k \omega
		-\nabla \cdot \left(\left(\mu + \sigma^\ast  \mu_t \right) \nabla k \right)
		\\
		&	\frac{\partial \rho \omega}{\partial t}
		+\nabla \cdot \left(\rho u \omega\right)
		=
		\frac{\gamma \rho \omega}{k} \tau_{t} \cddot \nabla u
		- \rho \beta \omega^2
		-\nabla \cdot \left( \left(\mu + \sigma \mu_t \right) \nabla \omega \right)
		+ \rho F  \frac{\sigma_\omega}{\omega} \nabla k \cdot \nabla \omega ,
	\end{aligned}
\end{equation}
where $\mu$ is the dynamic viscosity. The purpose of scaling Equation \ref{rhoKEqn} is to extend the turbulence model naturally to account for a system of two fluids. 
\subsection {Jump conditions at the fluid-fluid interface}
Applying the continuity of the interface (no mass transfer) yields a continuous velocity field across the fluid-fluid interface. 
\begin{equation}
	\lim\limits_{r\rightarrow0^+} u = \lim\limits_{r\rightarrow0^-} u
	\Rightarrow
	\lim\limits_{r\rightarrow0^+} \overline{u} = \lim\limits_{r\rightarrow0^-} \overline{u}
	,\text{ and}
	\quad
	\lim\limits_{r\rightarrow0^+} {u^\prime}
	=
	\lim\limits_{r\rightarrow0^-} {u^\prime},
\end{equation}
where $r$ is the radius of a ball centered at point $x \in \Gamma$. The positive sign indicates that the limit is taken from the side of the first region, and negative otherwise. $\overline{u}$ is the time-averaged velocity vector. Hence, the specific turbulent kinetic energy is also continuous across the interface. However, the convection of turbulent kinetic energy is not continuous as follows
\begin{equation}
	\lim\limits_{r\rightarrow0^+} k_1 = \lim\limits_{r\rightarrow0^-} k_2
	=
	\lim\limits_{r\rightarrow0} \frac{1}{2} 
	{u^\prime \cdot u^\prime}
	,\quad
	\llbracket \rho k \rrbracket 
	= 
	\llbracket \rho \rrbracket k.
\end{equation}
The instantaneous viscous dissipation rate and its associated jump read
\begin{equation}
	\epsilon = \nu 
	\left(
	\frac{\partial u_i}{\partial x_j}
	\right)^2
	\Rightarrow
	\llbracket \epsilon \rrbracket  
	= 
	\llbracket
	\nu 
	\left(
	\frac{\partial u_i}{\partial x_j}
	\right)^2
	\rrbracket
	.
	\label{epsilon_defn}
\end{equation}
\subsection{Dynamic problem of the interface}
Let the interface be described using a level set field $\psi$ whose unit normal vector pointing from the second to first phase is $n=\nabla \psi$. Moreover, the tangent space is formed by the tangent and bi-normal vectors $t$ and $b$ respectively. The tangent space and the unit normal vector form a complete basis on the interface which is assumed to have a unique orientation. The dynamics of the interface are governed by
\begin{equation}
	\frac{\partial \psi}{\partial t} +  u_1 \cdot n = 0
	,\quad
	-\frac{\partial \psi}{\partial t} -  u_2 \cdot n = 0
	\Rightarrow
	u_1 = u_2 = u ,\forall x \in \Gamma.
\end{equation}
Note that continuity of the velocity field does not imply a continuity in the associated derivatives. That means the interface does not alter the fact that the velocity field is continuous. Hence, for a two-phase flow problem
\begin{equation}
	u \in C^1.
\end{equation}
By differentiating the level set governing equation
\begin{equation}
	\nabla  \left(  \frac{\partial \psi}{\partial t}  \right) + 
	\nabla \left( u_1 \cdot n \right)
	=
	0
	.
\end{equation}
Expanding
\begin{equation}
	\nabla  \left(  \frac{\partial \psi}{\partial t}  \right) + 
	\left( \nabla u_1 \right) n 
	+
	\left( \nabla n \right) u_1 
	=
	0
	.
\end{equation}
Switching the order of the temporal and spatial derivatives reads
\begin{equation}
	\left(  \frac{\partial n}{\partial t}  \right)+ 
	\left( \nabla u_1 \right) n 
	+
	\left( \nabla n \right) u_1 
	=
	0
	.
\end{equation}
Similarly,
\begin{equation}
	-\left(  \frac{\partial n}{\partial t}  \right)+ 
	-\left( \nabla u_2 \right) n 
	-
	\left( \nabla n \right) u_2 
	=
	0
\end{equation}
Adding the two equations
\begin{equation}
	\left( \nabla u_1 \right) n 
	-
	\left( \nabla u_2 \right) n  
	=
	0
	.
\end{equation}
Then
\begin{equation}
	\left( \nabla u_1-\nabla u_2 \right) n 
	=
	0
	,
\end{equation}
and
\begin{equation}
	n \cdot \left(\nabla u \right) n
	=
	-
	\left(
	\frac{\partial n}{\partial t}
	+
	\left(\nabla n\right)u
	\right)
	\cdot 
	n
\end{equation}
\subsection{Kinematic problem of the interface}
The force balance in the normal direction is
\begin{equation}
	-p_1
	+ \frac{\mu_1}{2}  n \cdot
	\left( \nabla u_1 + \nabla u_1^T \right) n
	=
	-p_2 +  
	\frac{\mu_2}{2} 
	\cdot
	\left( \nabla u_2 + \nabla u_2^T \right) n.
\end{equation}
Note that
\begin{equation}
	n \cdot \left( \nabla u \right) n
	=
	n \cdot \left( \nabla u \right)^T n.
\end{equation}
Then
\begin{equation}
	-p_1
	+ \mu_1   
	n \cdot \left( \nabla u_1 \right) n
	=
	-p_2
	+ \mu_2   
	n \cdot \left( \nabla u_2 \right) n,
\end{equation}
and
\begin{equation}
	p_2 - p_1
	=
	\left( \mu_1 - \mu_2 \right)
	\left(
	\frac{\partial n}{\partial t}
	+
	\left(\nabla n\right)u
	\right)
	\cdot 
	n
\end{equation}
The force balance in the tangential direction is
\begin{equation}
	\frac{\mu_1}{2}  t \cdot \left(
	\left( \nabla u_1 + \nabla u_1^T \right) \right) n
	=
	\frac{\mu_2}{2}  t \cdot \left(
	\left( \nabla u_2 + \nabla u_2^T \right) \right) n
\end{equation}
The force balance in the bi-normal direction is
\begin{equation}
	\frac{\mu_1}{2}  b \cdot \left(
	\left( \nabla u_1 + \nabla u_1^T \right) \right) n
	=
	\frac{\mu_2}{2}  b \cdot \left(
	\left( \nabla u_2 + \nabla u_2^T \right) \right) n.
\end{equation}
\subsection{Viscous dissipation rate}
By writing the viscous dissipation rate in terms of $t,b,n$  (basis on the fluid-fluid interface), the dissipation rate reads
\begin{equation}
	\begin{aligned}
		\epsilon_1
		=
		\nu_1 \Bigl(
		&
		\left(\hat{t} \cdot \left(\nabla u_1\right)  \hat{t}\right)^2 +
		\left(\hat{t} \cdot \left(\nabla u_1\right)  \hat{b}\right)^2 +
		\left(\hat{t} \cdot \left(\nabla u_1\right)  \hat{n}\right)^2 + 
		\\
		&
		\left(\hat{b} \cdot \left(\nabla u_1\right)  \hat{t}\right)^2 +
		\left(\hat{b} \cdot \left(\nabla u_1\right)  \hat{b}\right)^2 +
		\left(\hat{b} \cdot \left(\nabla u_1\right)  \hat{n}\right)^2 + 
		\\
		&
		\left(\hat{n} \cdot \left(\nabla u_1\right)  \hat{t}\right)^2 +
		\left(\hat{n} \cdot \left(\nabla u_1\right)  \hat{b}\right)^2 +
		\left(\hat{n} \cdot \left(\nabla u_1\right)  \hat{n}\right)^2 
		\Bigl)
		,
	\end{aligned}	
\end{equation}
and
\begin{equation}
	\begin{aligned}
		\epsilon_2
		=
		\nu_2 \Bigl(
		&
		\left(\hat{t} \cdot \left(\nabla u_2\right)  \hat{t}\right)^2 +
		\left(\hat{t} \cdot \left(\nabla u_2\right)  \hat{b}\right)^2 +
		\left(\hat{t} \cdot \left(\nabla u_2\right)  \hat{n}\right)^2 + 
		\\
		&
		\left(\hat{b} \cdot \left(\nabla u_2\right)  \hat{t}\right)^2 +
		\left(\hat{b} \cdot \left(\nabla u_2\right)  \hat{b}\right)^2 +
		\left(\hat{b} \cdot \left(\nabla u_2\right)  \hat{n}\right)^2 + 
		\\
		&
		\left(\hat{n} \cdot \left(\nabla u_2\right)  \hat{t}\right)^2 +
		\left(\hat{n} \cdot \left(\nabla u_2\right)  \hat{b}\right)^2 +
		\left(\hat{n} \cdot \left(\nabla u_2\right)  \hat{n}\right)^2 
		\Bigl)
		.
	\end{aligned}	
\end{equation}
Substitutions for the three force components
\begin{equation}
	t \cdot \left(\frac{\mu_2}{2} 
	\left( \nabla u_2 + \nabla u_2^T \right) \right) n
	=
	\frac{\mu_2}{2} 
	\left( 
	t \cdot \left( \nabla u_2 \right) n
	+
	n \cdot \left( \nabla u_2 \right) t
	\right).
\end{equation}
This is obtained from
\begin{equation}
	m \cdot A n = n \cdot A^T m.
\end{equation}
From the dynamic condition
\begin{equation}
	t \cdot \left( \nabla u_1 \right) n = t \cdot \left( \nabla u_2 \right) n.
\end{equation}
Substitution
\begin{equation}
	n \cdot \left( \nabla u_2 \right) t =
	\frac{\mu_1 -\mu_2}{\mu_2} t \cdot \left( \nabla u_1 \right) n 
	+ 
	\frac{\mu_1}{\mu_2} n \cdot \left( \nabla u_1 \right) t .
\end{equation}
Similarly,
\begin{equation}
	n \cdot \left( \nabla u_2 \right) b =
	\frac{\mu_1 -\mu_2}{\mu_2} b \cdot \left( \nabla u_1 \right) n 
	+ 
	\frac{\mu_1}{\mu_2} n \cdot \left( \nabla u_1 \right) b .
\end{equation}
From the dynamic condition
\begin{equation}
	n \cdot \left( \nabla u_2 \right) n = n \cdot \left( \nabla u_1 \right) n , \quad
	t \cdot \left( \nabla u_2 \right) n = t \cdot \left( \nabla u_1 \right) n , \quad
	b \cdot \left( \nabla u_2 \right) n = b \cdot \left( \nabla u_1 \right) n .
\end{equation}
Then, it remains $4$ terms in each fluid which do not require any jump to satisfy the stress balance
\begin{equation}
	t \cdot \left( \nabla u_{1,2} \right) t , \quad
	t \cdot \left( \nabla u_{1,2} \right) b , \quad
	b \cdot \left( \nabla u_{1,2} \right) t, \quad
	b \cdot \left( \nabla u_{1,2} \right) b.
\end{equation}
The effect of each term on the jump of the dissipation rate is as follows
\begin{equation}
	\nu_1 \left(\hat{t} \cdot \left(\nabla u_1\right)  \hat{n}\right)^2
	-
	\nu_2 \left(\hat{t} \cdot \left(\nabla u_2\right)  \hat{n}\right)^2
	=
	\left( \nu_1 - \nu_2 \right) \left(\hat{t} \cdot \left(\nabla u_1\right)  \hat{n}\right)^2.
\end{equation}
\begin{equation}
	\nu_1 \left(\hat{b} \cdot\left(\nabla u_1\right)  \hat{n}\right)^2
	-
	\nu_2 \left(\hat{b} \cdot\left(\nabla u_2\right)  \hat{n}\right)^2
	=
	\left( \nu_1 - \nu_2 \right) \left(\hat{b} \cdot \left(\nabla u_1\right)  \hat{n}\right)^2.
\end{equation}
\begin{equation}
	\nu_1 \left(\hat{n} \cdot \left(\nabla u_1\right)  \hat{n}\right)^2
	-
	\nu_2 \left(\hat{n} \cdot \left(\nabla u_2\right)  \hat{n}\right)^2
	=
	\left( \nu_1 - \nu_2 \right) \left(\hat{n} \cdot \left(\nabla u_1\right)  \hat{n}\right)^2.
\end{equation}
By substitution,
\begin{equation}
	\begin{aligned}
		\nu_1 \left(\hat{n} \cdot \left(\nabla u_1\right)  \hat{t}\right)^2
		-
		&\nu_2 \left(\hat{n} \cdot \left(\nabla u_2\right)  \hat{t}\right)^2
		=
		\nu_1 \left(\hat{n} \cdot \left(\nabla u_1\right)  \hat{t}\right)^2
		-
		\nu_2 \left( \frac{\mu_1 -\mu_2}{\mu_2} t \cdot \left( \nabla u_1 \right) n + \frac{\mu_1}{\mu_2} n \cdot \left( \nabla u_1 \right) t  \right)^2
		\\
		&
		=
		\nu_1 \left(\hat{n} \cdot \left(\nabla u_1\right)  \hat{t}\right)^2
		-
		\nu_2 \left( \frac{\mu_1 -\mu_2}{\mu_2} \right)^2  \left( t \cdot \left( \nabla u_1 \right) n \right)^2
		- \nu_2 \left(  \frac{\mu_1}{\mu_2}\right)^2  \left( n \cdot \left( \nabla u_1 \right) t \right)^2\\
		&
		-2 \nu_2 \left( \frac{\mu_1 -\mu_2}{\mu_2} \right) \left(  \frac{\mu_1}{\mu_2}\right) \left( t \cdot \left( \nabla u_1 \right) n \right) \left( n \cdot \left( \nabla u_1 \right) t \right)\\
		&=
		\left( \nu_1-\nu_2  \left(  \frac{\mu_1}{\mu_2}\right)^2  \right) \left( n \cdot \left( \nabla u_1 \right) t \right)^2
		- \nu_2 \left( \frac{\mu_1 -\mu_2}{\mu_2} \right)^2  \left( t \cdot \left( \nabla u_1 \right) n \right)^2\\
		&-2 \nu_2 \left( \frac{\mu_1 -\mu_2}{\mu_2} \right) \left(  \frac{\mu_1}{\mu_2}\right) \left( t \cdot \left( \nabla u_1 \right) n \right) \left( n \cdot \left( \nabla u_1 \right) t \right).
	\end{aligned}
\end{equation}
Similarly,
\begin{equation}
	\begin{aligned}
		\nu_1 \left(\hat{n} \cdot \left(\nabla u_1\right)  \hat{b}\right)^2
		-
		\nu_2 \left(\hat{n} \cdot \left(\nabla u_2\right)  \hat{b}\right)^2
		&=
		\left( \nu_1-\nu_2  \left(  \frac{\mu_1}{\mu_2}\right)^2  \right) \left( n \cdot \left( \nabla u_1 \right) b \right)^2
		- \nu_2 \left( \frac{\mu_1 -\mu_2}{\mu_2} \right)^2  \left( b \cdot \left( \nabla u_1 \right) n \right)^2\\
		&-2 \nu_2 \left( \frac{\mu_1 -\mu_2}{\mu_2} \right) \left(  \frac{\mu_1}{\mu_2}\right) \left( b \cdot \left( \nabla u_1 \right) n \right) \left( n \cdot \left( \nabla u_1 \right) b \right).
	\end{aligned}
\end{equation}
Then the jump due to the previous $5$ terms read
\begin{equation}
	\begin{aligned}
		q_1-q_2
		&=
		\left( \nu_1 - \nu_2 \right) \left(\hat{n} \cdot \left(\nabla u_1\right)  \hat{n}\right)^2
		\\&
		+
		\left( \nu_1 - \nu_2 - \nu_2 \left( \frac{\mu_1 -\mu_2}{\mu_2} \right)^2 \right)
		\left( 
		\left(\hat{t} \cdot \left(\nabla u_1\right)  \hat{n}\right)^2
		+
		\left(\hat{b} \cdot \left(\nabla u_1\right)  \hat{n}\right)^2
		\right)
		\\
		&
		+
		\left( \nu_1-\nu_2  \left(  \frac{\mu_1}{\mu_2}\right)^2  \right) 
		\left(
		\left( n \cdot \left( \nabla u_1 \right) t \right)^2
		+
		\left( n \cdot \left( \nabla u_1 \right) b \right)^2
		\right)
		\\
		&
		-2 \nu_2 \left( \frac{\mu_1 -\mu_2}{\mu_2} \right) \left(  \frac{\mu_1}{\mu_2}\right) 
		\left(
		\left( b \cdot \left( \nabla u_1 \right) n \right) \left( n \cdot \left( \nabla u_1 \right) b \right)
		+
		\left( t \cdot \left( \nabla u_1 \right) n \right) \left( n \cdot \left( \nabla u_1 \right) t \right)
		\right).
	\end{aligned}
\end{equation}
The jump in the rest of the dissipation terms which are not related to any jump conditions are
\begin{equation}
	\begin{aligned}
		h_1-h_2
		&=
		\nu_1 \left( 
		\left(\hat{t} \cdot  \left(\nabla u_1\right)  \hat{t}\right)^2
		+
		\left(\hat{t} \cdot  \left(\nabla u_1\right)  \hat{b}\right)^2
		+
		\left(\hat{b} \cdot  \left(\nabla u_1\right)  \hat{t}\right)^2
		+
		\left(\hat{b} \cdot  \left(\nabla u_1\right)  \hat{b}\right)^2
		\right)
		\\&
		-
		\nu_2 \left( 
		\left(\hat{t} \cdot  \left(\nabla u_2\right)  \hat{t}\right)^2
		+
		\left(\hat{t} \cdot  \left(\nabla u_2\right)  \hat{b}\right)^2
		+
		\left(\hat{b} \cdot  \left(\nabla u_2\right)  \hat{t}\right)^2
		+
		\left(\hat{b} \cdot  \left(\nabla u_2\right)  \hat{b}\right)^2
		\right)
		\\
		& =
		\left( \nu_1 - \nu_2 \right)
		\left( 
		\left(\hat{t} \cdot  \left(\nabla u_1\right)  \hat{t}\right)^2
		+
		\left(\hat{t} \cdot  \left(\nabla u_1\right)  \hat{b}\right)^2
		+
		\left(\hat{b} \cdot  \left(\nabla u_1\right)  \hat{t}\right)^2
		+
		\left(\hat{b} \cdot  \left(\nabla u_1\right)  \hat{b}\right)^2
		\right)
		.
	\end{aligned}
\end{equation}
The total jump in the dissipation rate reads
\begin{equation}
	\begin{aligned}
		\epsilon_1-\epsilon_2
		&=
		\left( \nu_1 - \nu_2 \right)
		\left( 
		\left(\hat{n} \cdot \left(\nabla u_1\right)  \hat{n}\right)^2
		+
		\left(\hat{t} \cdot \left(\nabla u_1\right)  \hat{t}\right)^2
		+
		\left(\hat{t} \cdot \left(\nabla u_1\right)  \hat{b}\right)^2
		+
		\left(\hat{b} \cdot \left(\nabla u_1\right)  \hat{t}\right)^2
		+
		\left(\hat{b} \cdot \left(\nabla u_1\right)  \hat{b}\right)^2
		\right)
		\\&
		+
		\left( \nu_1 - \nu_2 - \nu_2 \left( \frac{\mu_1 -\mu_2}{\mu_2} \right)^2 \right)
		\left( 
		\left(\hat{t} \cdot \left(\nabla u_1\right)  \hat{n}\right)^2
		+
		\left(\hat{b} \cdot \left(\nabla u_1\right)  \hat{n}\right)^2
		\right)
		\\
		&
		+
		\left( \nu_1-\nu_2  \left(  \frac{\mu_1}{\mu_2}\right)^2  \right) 
		\left(
		\left( n \cdot \left( \nabla u_1 \right) t \right)^2
		+
		\left( n \cdot \left( \nabla u_1 \right) b \right)^2
		\right)
		\\
		&
		+2 \nu_2 \left( \frac{\mu_2 -\mu_1}{\mu_2} \right) \left(  \frac{\mu_1}{\mu_2}\right) 
		\left(
		\left( b \cdot \left( \nabla u_1 \right) n \right) \left( n \cdot \left( \nabla u_1 \right) b \right)
		+
		\left( t \cdot \left( \nabla u_1 \right) n \right) \left( n \cdot \left( \nabla u_1 \right) t \right)
		\right)
		.
	\end{aligned}
	\label{exactJump}
\end{equation}
Note: Equation \ref{exactJump} is exact and does not contain any assumptions.
\newline
Further simplific ations are
\begin{equation}
	\nu_1 - \nu_2 - \nu_2 \left( \frac{\mu_1 -\mu_2}{\mu_2} \right)^2 
	=
	\left( \nu_1 - \nu_2\right)
	-
	\nu_2 \left( \frac{\mu_1 -\mu_2}{\mu_2} \right)^2,
	\quad
	\nu_1-\nu_2  \left(  \frac{\mu_1}{\mu_2}\right)^2  =
	\left( \nu_1 - \nu_2\right)
	+
	\nu_2  \left( 1 - \left( \frac{\mu_1}{\mu_2} \right)^2 \right).
\end{equation}
For the case of air and water
\begin{equation}
	\nu_1 = 1e^{-5}, \quad \mu_1 = 1e^{-5}, \quad \nu_2 = 1e^{-6} \quad \text{and \hspace{0.5cm}} \mu_2 = 1e^{-3}.
\end{equation}
Then
\begin{equation}
	\nu_2 \left( \frac{\mu_1 -\mu_2}{\mu_2} \right)^2 \sim \nu_2, \quad
	\nu_2  \left( 1 - \left( \frac{\mu_1}{\mu_2} \right)^2 \right) \sim \nu_2.
\end{equation}
The jump for such fluids reads
\begin{equation}
	\begin{aligned}
		\epsilon_1-\epsilon_2
		&=
		\frac{\left( \nu_1 - \nu_2 \right)}{\nu_1}
		\epsilon_1
		\\&
		+
		\nu_2
		\left( 
		\left(\hat{n} \cdot \left(\nabla u_1\right)  \hat{t}\right)^2
		+
		\left(\hat{n} \cdot \left(\nabla u_1\right)  \hat{b}\right)^2
		-
		\left(\hat{t} \cdot \left(\nabla u_1\right)  \hat{n}\right)^2
		-
		\left(\hat{b} \cdot \left(\nabla u_1\right)  \hat{n}\right)^2
		\right)
		\\&
		+2 \nu_2 \left( \frac{\mu_2 -\mu_1}{\mu_2} \right) \left(  \frac{\mu_1}{\mu_2}\right) 
		\left(
		\left( b \cdot \left( \nabla u_1 \right) n \right) \left( n \cdot \left( \nabla u_1 \right) b \right)
		+
		\left( t \cdot \left( \nabla u_1 \right) n \right) \left( n \cdot \left( \nabla u_1 \right) t \right)
		\right)
		.
	\end{aligned}
\end{equation}
The jump is further scaled wirth respect to the first fluid as follows
\begin{equation}
	\begin{aligned}
		\frac{\epsilon_1-\epsilon_2}
		{\frac{\left( \nu_1 - \nu_2 \right)}{\nu_1}
			\epsilon_1}
		&=
		1
		\\&
		+
		\frac{\nu_2}{\nu_1 - \nu_2}
		\frac{
			\nu_1
			\left( 
			\left(\hat{n} \cdot \left(\nabla u_1\right)  \hat{t}\right)^2
			+
			\left(\hat{n} \cdot \left(\nabla u_1\right)  \hat{b}\right)^2
			-
			\left(\hat{t} \cdot \left(\nabla u_1\right)  \hat{n}\right)^2
			-
			\left(\hat{b} \cdot \left(\nabla u_1\right)  \hat{n}\right)^2
			\right)
		}
		{
			\epsilon_1
		}
		\\&
		+
		\frac{2 \nu_2 \left( \frac{\mu_2 -\mu_1}{\mu_2} \right) \left(  \frac{\mu_1}{\mu_2}\right) }{\nu_1 - \nu_2}
		\frac{\nu_1 \left(
			\left( b \cdot \left( \nabla u_1 \right) n \right) \left( n \cdot \left( \nabla u_1 \right) b \right)
			+
			\left( t \cdot \left( \nabla u_1 \right) n \right) \left( n \cdot \left( \nabla u_1 \right) t \right)
			\right)}
		{\epsilon_1}
		.
	\end{aligned}
\end{equation}
The order of the second term of the right hand side is
\begin{equation}
	\norm{\frac{\nu_2}{\nu_1 - \nu_2}
		\frac{
			\nu_1
			\left( 
			\left(\hat{n} \cdot \left(\nabla u_1\right)  \hat{t}\right)^2
			+
			\left(\hat{n} \cdot \left(\nabla u_1\right)  \hat{b}\right)^2
			-
			\left(\hat{t} \cdot \left(\nabla u_1\right)  \hat{n}\right)^2
			-
			\left(\hat{b} \cdot \left(\nabla u_1\right)  \hat{n}\right)^2
			\right)
		}
		{
			\epsilon_1
		}
	}
	\leq
	\frac{\nu_2}{\nu_1 - \nu_2}
	\leq
	0.1,
\end{equation}
for the system of air-water.
\newline
Similarly, the third term is bounded as follows
\begin{equation}
	\begin{aligned}
		\norm{
			\frac{2 \nu_2 \left( \frac{\mu_2 -\mu_1}{\mu_2} \right) \left(  \frac{\mu_1}{\mu_2}\right) }{\nu_1 - \nu_2}
			\frac{\nu_1 \left(
				\left( b \cdot \left( \nabla u_1 \right) n \right) \left( n \cdot \left( \nabla u_1 \right) b \right)
				+
				\left( t \cdot \left( \nabla u_1 \right) n \right) \left( n \cdot \left( \nabla u_1 \right) t \right)
				\right)}
			{\epsilon_1}
		}
		&\leq
		\norm{
			\frac{2 \nu_2 \left( \frac{\mu_2 -\mu_1}{\mu_2} \right) \left(  \frac{\mu_1}{\mu_2}\right) }{\nu_1 - \nu_2}
		}\\
		&\leq
		0.2 \frac{10^{-5}}{10^{-3}}
		\leq
		2e-3,
	\end{aligned}
\end{equation}
for the system of air-water.
\newline
The linearised (decomposed) jump of the dissipation relative to the first fluid is bounded as follows
\begin{equation}
	0.9
	\leq
	\norm{\frac{\epsilon_1-\epsilon_2}
		{\frac{\left( \nu_1 - \nu_2 \right)}{\nu_1}
			\epsilon_1}}
	\leq
	1.1.
\end{equation}
Since the dissipation rate is always positive, then
\begin{equation}
	\frac{\epsilon_2}{\epsilon_1} \leq \frac{\nu_1}{\nu_1-\nu_2} \theta^2, \quad \text{for system air-water }\theta^2=0.1, \frac{\nu_1}{\nu_1-\nu_2}=\frac{1}{0.9},  	\frac{\epsilon_2}{\epsilon_1} \leq 0.111 \lessapprox 0.1.
\end{equation}
Consider decomposing the dissipation into mean and perturbation as well
\begin{equation}
	\epsilon = \Bar{\epsilon}+\epsilon^\prime.
\end{equation}
Then
\begin{equation}
	\epsilon_2 \leq \theta^2 \epsilon_1 \Rightarrow \Bar{\epsilon_2} \leq \theta^2 \Bar{\epsilon_1},
\end{equation}
and
\begin{equation}
	\Bar{\epsilon_2}+\epsilon_2^\prime \leq \theta^2 \Bar{\epsilon_1}+\theta^2 \epsilon_1^\prime, \quad
	\epsilon_2^\prime \leq \theta^2 \Bar{\epsilon_1} - \Bar{\epsilon_2} +\theta^2 \epsilon_1^\prime.
\end{equation}
By normalising
\begin{equation}
	\frac {\epsilon_2^\prime}{\epsilon_1^\prime} \leq 
	\theta^2 \frac{\Bar{\epsilon_1}}{\epsilon_1^\prime} - \frac{\Bar{\epsilon_2}}{\epsilon_1^\prime} +\theta^2 \leq
	\theta^2 \frac{\Bar{\epsilon_1}}{\epsilon_1^\prime} - 
	\frac{\Bar{\epsilon_2}}{\Bar{\epsilon_1}}\frac{\Bar{\epsilon_1}}{\epsilon_1^\prime}
	+\theta^2
	.
\end{equation}
Considering the two cases (bounds) of the inequality of the ratio of the mean as follows: (1) first case:
\begin{equation}
	\frac {\epsilon_2^\prime}{\epsilon_1^\prime} \leq 
	\theta^2 \frac{\Bar{\epsilon_1}}{\epsilon_1^\prime} - 
	\theta^2\frac{\Bar{\epsilon_1}}{\epsilon_1^\prime}
	+\theta^2
	\leq \theta^2
	,
\end{equation}
and (2) second case:
\begin{equation}
	\frac {\epsilon_2^\prime}{\epsilon_1^\prime} \leq 
	\theta^2 \frac{\Bar{\epsilon_1}}{\epsilon_1^\prime} 
	+\theta^2.
\end{equation}
By employing order of magnitude approximation, let the mean velocity scale be $\Bar{V}$, perturbation $v^\prime$, turbulent intensity $I=\left(\frac{v^\prime}{\Bar{V}}\right)^2$, laminar length scale $L$ and turbulent length scale $l$. The mean dissipation rate scales as $\Bar{\epsilon} \sim \frac{\Bar{V}^2}{L^2}$ and the perturbation scales as ${\epsilon^\prime} \sim \frac{{v^\prime}^2}{l^2}$. Then
\begin{equation}
	\frac{\Bar{\epsilon}}{\epsilon^\prime} 
	\sim
	\left(\frac{\Bar{V}}{v^\prime}\right)^2
	\left(\frac{l}{L}\right)^2
	\sim
	\frac{1}{I}
	\left(\frac{l}{L}\right)^2.
\end{equation}
For moderate turbulent intensities $I\sim 0.1$ and it assuming that the ratio between the turbulent length scale to the laminar (mean flow) length scale is much less than unity (significantly small) $\frac{l}{L}<1 \Rightarrow \left(\frac{l}{L}\right)^2 \ll 1$. Therefore
\begin{equation}
	\theta^2 \frac{\Bar{\epsilon_1}}{\epsilon_1^\prime}  < \theta^2.
\end{equation}
Under theses conditions the normalised perturbation in the dissipation rate is bounded as follows
\begin{equation}
	\frac {\epsilon_2^\prime}{\epsilon_1^\prime} \leq 
	2\theta^2.
\end{equation}
For the system of air-water $\theta^2=0.1$. Then,
\begin{equation}
	\frac {\epsilon_2^\prime}{\epsilon_1^\prime} \leq 0.2.
\end{equation}

The simplified jump condition of the instantaneous viscous dissipation rate assuming continuous velocity derivatives reads 
\begin{equation}
	\llbracket \epsilon \rrbracket  
	\simeq
	\llbracket \nu \rrbracket
	\left(
	\frac{\partial u_i}{\partial x_j}
	\right)^2
	,
\end{equation}
which computes the jump with an error $\pm 10\%$ for the system of water and air. The simplified jump considers only the effect of the viscosity and neglects the jump in the velocity derivatives (assumed to be continuous).
\newline
Under this assumption (simplification) the velocity field is first order continuous. Accordingly, both the mean and perturbation fields have at least continuous first order derivatives.
\begin{equation}
	\left(
	\frac{\partial u_i}{\partial x_j}
	\right)^2 \in C^0, \Rightarrow
	u \in C^1, 
	\quad 
	\frac{\partial u^{\prime}_{i}}{\partial x_j} \in C^0,
	\frac{\partial \Bar{u}_{i}}{\partial x_j} \in C^0.
\end{equation}
\newline
Accordingly, the specific dissipation rate (frequency) follows the same jump
\begin{equation}\label{jumpOmega}
	\omega = \epsilon_{turb} / k \Rightarrow
	\llbracket \omega \rrbracket 
	= 
	\llbracket \epsilon_{turb} \rrbracket
	/ k
	=
	\llbracket \nu \rrbracket
	\left(
	\frac{\partial u^\prime_i}{\partial x_j}
	\right)^2
	\frac{1}{k}
	, \quad
	\llbracket \rho \omega \rrbracket 
	=
	\llbracket \mu \rrbracket
	\left(
	\frac{\partial u^\prime_i}{\partial x_j}
	\right)^2
	\frac{1}{k}
	.
\end{equation}
The jump in the gradient of the turbulence dissipation frequency is
\begin{equation}\label{jumpOmegaGradient}
	\llbracket \nabla \omega \rrbracket
	=
	\llbracket \nu \rrbracket
	\nabla
	\left(
	\frac{1}{k}
	\left(
	\frac{\partial u^\prime_i}{\partial x_j}
	\right)^2
	\right)
	.
\end{equation}
The jump in the turbulent kinematic and dynamic viscosity read
\begin{equation}
	\llbracket \nu_t \rrbracket 
	= \frac{k}{\omega_1} - \frac{k}{\omega_2}
	= k \llbracket \frac{1}{\omega}\rrbracket 
	= \frac{k^2}{\left(
		\frac{\partial u^\prime_i}{\partial x_j}
		\right)^2} 
	\llbracket \frac{1}{\nu} \rrbracket
	,
	\quad
	\llbracket \mu_t \rrbracket 
	= \frac{k}{\left(
		\frac{\partial u^\prime_i}{\partial x_j}
		\right)^2} 
	\llbracket \frac{1}{\mu} \rrbracket
	.
\end{equation}
similarly
\begin{equation}
	\llbracket \mu \nabla \omega \rrbracket =
	{\llbracket \mu \nu \rrbracket}
	\nabla
	\left(
	\frac{1}{k}
	\left(
	\frac{\partial u^\prime_i}{\partial x_j}
	\right)^2 
	\right)
	,
	\quad
	\llbracket \nu \nabla \omega \rrbracket =
	{\llbracket \nu^2 \rrbracket}
	\nabla
	\left(
	\frac{1}{k}
	\left(
	\frac{\partial u^\prime_i}{\partial x_j}
	\right)^2 
	\right).
\end{equation}

\subsection{The inverse turbulence area}
The inverse turbulent area field is obtained by dividing the turbulent dissipation frequency by the kinematic viscosity and assuming negligible effect of the velocity gradient jump condition. It has units of $\text{ m}^{-2}$, which suggests the name of inverse turbulence area.
\begin{equation}
	\label{defn_zeta}
	\zeta=\frac{\omega}{\nu} = 
	\frac{1}{k}
	\left(
	\frac{\partial u^\prime_i}{\partial x_j}
	\right)^2,
	\quad
	[\zeta]=m^{-2},
\end{equation}
which is governed by the following transport equation
\begin{equation}\label{zetaEqn}
	\frac{\partial \zeta}{\partial t}
	+
	\nabla \cdot \left(u \zeta\right)
	=
	\frac{\gamma \zeta}{k} 
	- \beta \nu \zeta^2
	-\nabla \cdot \left( \nu_{eff} \nabla \zeta \right)
	+F  \frac{\sigma_\omega}{ \nu \omega} \nabla k \cdot \nabla \omega
	,
\end{equation}
or in the integral form as follows
\begin{equation}\label{integralZetaEqnApproximated}
	\frac{\partial}{\partial t}
	\int\limits_{\Omega} \rho \zeta dV
	+\int\limits_{\partial \Omega} 
	\rho u \zeta \cdot dS
	=
	\int\limits_{\Omega} \frac{\gamma \zeta}{k}  dV
	- \int\limits_{\Omega} \beta \nu \zeta^2
	-	\int\limits_{\partial \Omega}   \nu_{eff} \nabla \zeta \cdot dS
	+\int\limits_{\Omega}   \frac{F \sigma_\omega}{\nu \omega} \nabla k \cdot \nabla \omega dV
	.
\end{equation}
\subsection{Remarks on the system of air-water waves}
In the preceding derivations, the buoyancy modification source term (see for example \cite{devolder,larsen_fuhrman_2018}) is not used since it does not appear naturally in the equations. The presented model does not consider the dynamics of dispersive flows, where buoyancy force exists. This term laminarize the flow conditions around the interface in both regions in the vicinity of the interface.
\newline
Furthermore, the dynamic viscosity is $10^{-5} \text{ m}^2/\text{s}$ for the air and is $10^{-6} \text{ m}^2/\text{s}$ for the water. From Equation \ref{jumpOmega}, the dissipation frequency in the air is $10$ times its value in the water for the same turbulence kinetic energy. Hence, turbulent eddies remain approximately $0.1$ times their life-time in air in comparison to the water (one order of magnitude). The previous finding explains the low turbulence level in the water for air-water waves, despite the fact that it might grow at larger values in the air.

\section{Numerical approximation}
\label{sec:numerics}
The finite volume method (FVM) is used for the spatial discretization and first order finite difference is employed for temporal discretization. The integrals over the fluid-fluid interface are assumed to be negligible for sufficiently fine grids. Consider a finite volume (cell $P$) with volume $V_P$ and an $n$ flat faces. Each face $f$ has a neighbor cell $N\left(f\right)$ and an area vector $S_f$ pointing outward of $P$. The interpolation is performed linearly based on the distance yielding two weights $\psi$ and $\left(1-\psi\right)$ for the owner and neighbor cell respectively. The velocity flux at each face is $u_f \cdot S_f=\phi_f$. The superscript $\left(^0\right)$ represents the values of the field at the previous time-step. The discrete approximation of the inverse turbulence area (Equation \ref{integralZetaEqnApproximated}) reads
\begin{equation}
\begin{aligned}
	\label{zeta_discretized}
	&
	\rho_P \zeta_P- \rho_P^0 \zeta_P^0
	+
	\frac{1}{V_P \Delta t}\sum\limits_{f} 
	\left(\psi_f \zeta_P  + \left(1-\psi_f\right) \zeta_{N\left(f\right)}\right) \rho_f \phi_f
	=\frac{\rho_P}{ \Delta t} \frac{\gamma}{k_P} \zeta_P \\
	&-
	\frac{\rho_P}{ \Delta t} \beta \nu _P \zeta_P^0 \zeta_P-
	\frac{1}{V_P \Delta t}\sum\limits_{f} \mu_{eff,f} 
	\frac{\zeta_P-\zeta_{N\left(f\right)}}
	 { 
	 	\lVert {x_P-x_{N\left(f\right)}} \rVert
	 }
	\lVert{S_f}\rVert
	+
	F \frac{\sigma_\omega}{\nu_P \omega_P^0} \nabla k_P^0 \cdot \nabla \omega_P^0
	.
\end{aligned}
\end{equation}
The discretized approximation of the turbulent kinetic energy (Equation \ref{integralKEqn}) reads
\begin{equation}
	\begin{aligned}
		\label{k_discretized}
		\rho_P k_P- \rho_P^0 k_P^0
		+
		\frac{1}{V_P \Delta t}\sum\limits_{f} 
		&\left(\psi_f k_P  + \left(1-\psi_f\right) k_{N\left(f\right)}\right) \rho_f \phi_f
		=\\
		\frac{\rho_P}{ \Delta t} \tau_{t,P}^0 \cddot \nabla u -
		&\frac{\rho_P}{ \Delta t} \beta^\ast \omega_P k_P-
		\frac{1}{V_P \Delta t}\sum\limits_{f} \mu_{eff,f} 
		\frac{k_P-k_{N\left(f\right)}}
		{ 
			\lVert {x_P-x_{N\left(f\right)}} \rVert
		}
		\lVert{S_f}\rVert
		.
	\end{aligned}
\end{equation}
The dissipation frequency and turbulent kinematic viscosity read
\begin{equation}
	\omega_p = \zeta_P\nu_P, \quad \nu_{t,P} = \frac{k_P}{\nu_P \zeta_P} L\left(\omega_p\right),
\end{equation}
where $L\left(\omega_p\right)$ is a limiter function to bound the turbulent stress or stabilize the model (\cite{larsen_fuhrman_2018}). Moreover, the model parameters correspond to the $k\omega SST$ model \cite{menter_3}.
\section{Application to Breaking waves}
\label{sec:results}
\begin{figure}
	\scalebox{0.2}{
		\includegraphics[trim={0cm 10cm 0cm 10cm},clip] {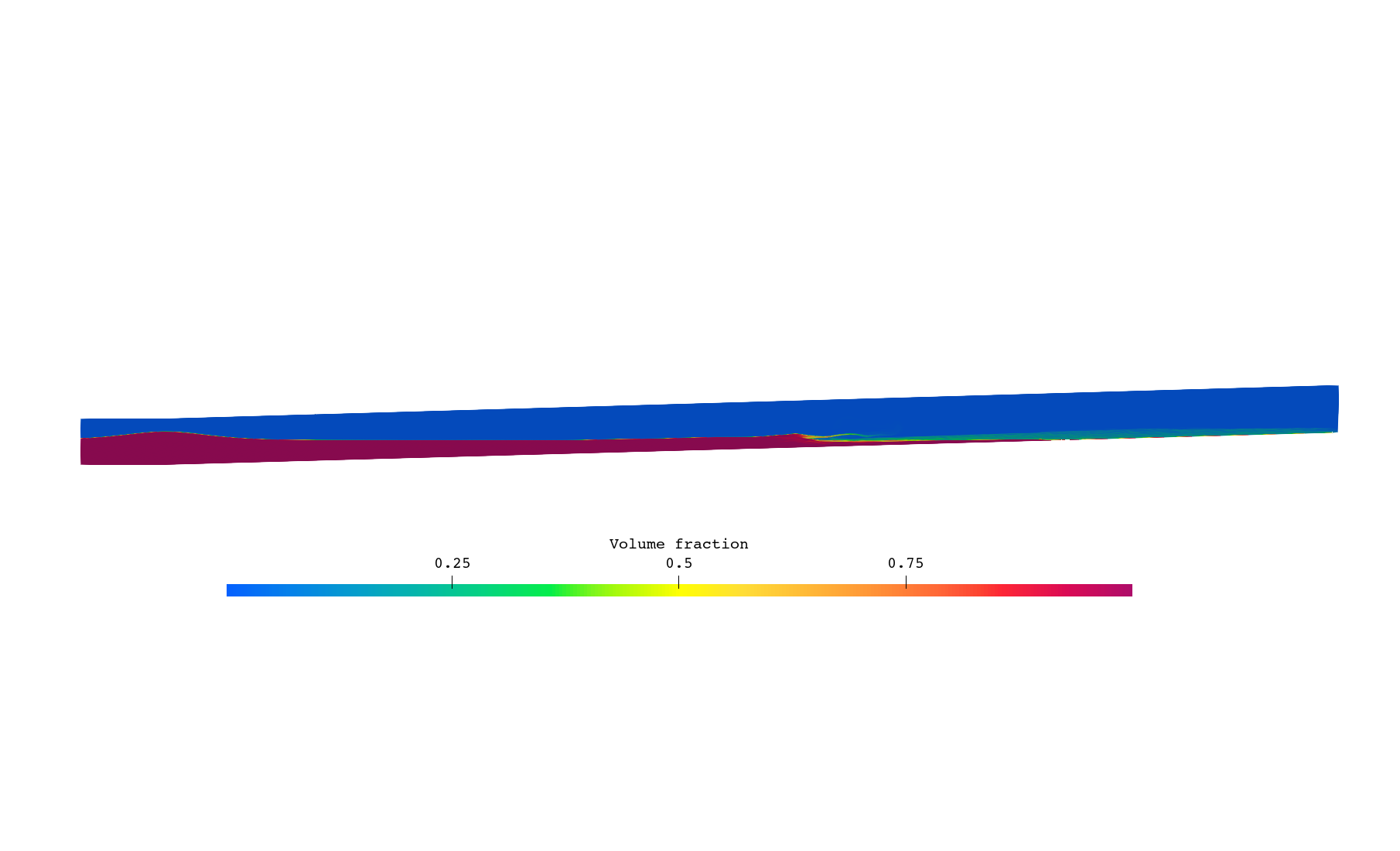}
	}
	\caption{Computational domain and volume fraction distribution at end of simulation.}
	\label{cmp_domain}
\end{figure}

A two-dimensional setup of the computational domain as reported in \cite{devolder} is used. The entrance (left side) is a rectangle with width of $1.3$ m and height $0.7$ m. The beach is modeled as sloped line. The slope is $1 \cddot 35$. The horizontal distance is $17.5$ m. The vertical distance is kept the same as in the inlet $0.7$ m as shown in Figure \ref{cmp_domain}. The $x-$ axis has value of $-2$ m at the inlet.
\newline
The computational grid is constructed using blockMesh utility in order to construct a multi-block structured grids. The number of quads is $315$ thousands elements. The size of each quad is $0.018$ m and $0.0028$ m in the horizontal and vertical direction, respectively.
\newline
Two wave conditions were considered in the present simulations. The wave depth is $0.4$ m. The first wave condition is spilling breaker type. The wave height is $0.125$ m and the wave period is $2$ s. The total simulation time is $100$ s or $50$ wave periods. The number of warm-up cycles is $30$. Only the last $20$ cycles are considered for the presented results. The second wave condition is plunging breaker type. The wave height is $0.127$ m and the wave period is $5$ s. The total simulation time is $250$ s or $50$ wave periods as well. Similarly, the number of warm-up cycles is $30$. Only the last $20$ cycles are considered for the presented results.
\newline
The domain is initialized with calm water. The wave is generated at the inlet via a wave generation boundary condition for both the volume fraction $\alpha$ and the velocity $u$. The pressure boundary condition is set to Neumann boundary condition. The turbulence kinetic energy is set to $10^{-6} \text{ m}^2/\text{s}^2$. The inverse turbulence area is set to $10^{6} \text{ m}^-2$ which corresponds to turbulence frequency of $1 \text{ s}^{-1}$ in the water similar to \cite{devolder}. At the wall, the velocity is set to no-slip boundary condition and everything else is set to Neumann. At the outlet and the atmosphere (upper lines) , the pressure is set to $0 \text{ Pa}$ and everything else is set to Neumann.
\newline
Figures \ref{vel_spilling} and \ref{vel_plunging} show the mean horizontal velocity ($u_x$) on the horizontal axis and the vertical distance from the bottom surface of the domain on the vertical axis. The experimental measurements show that the velocity variation across the vertical line is minimum. The mean bulk velocity is in the negative $x-$axis. The direction of the flow flips above the air-water interface. The velocity profiles associated with the presented turbulence model agree with the experimental data. The derivative of the horizontal velocity near the wall is significantly less than the case of $k \omega SST$ (\cite{devolder}). In general, it can be noticed that the presented model has a better agreement especially in case of the plunging breakers.
\newline
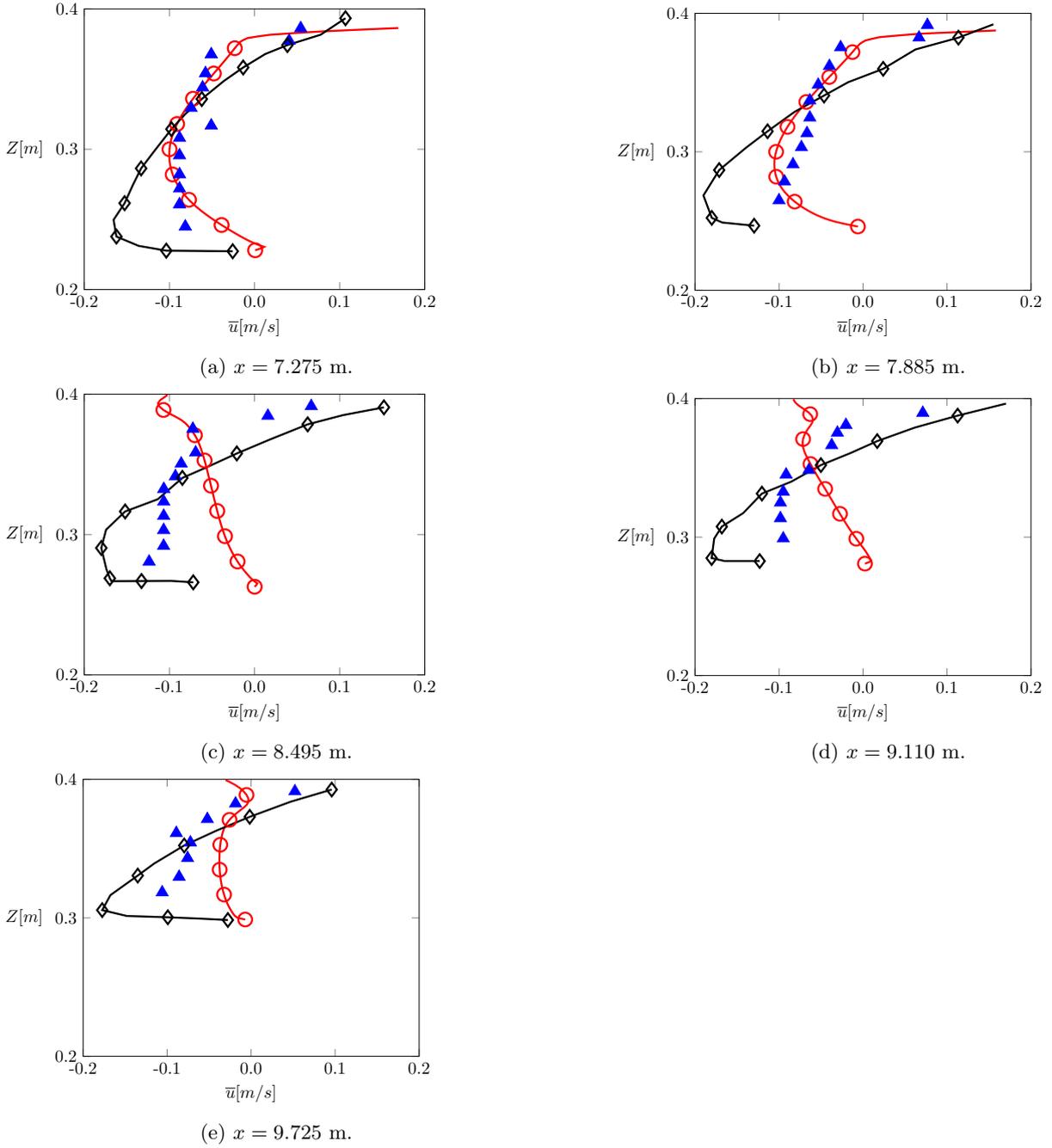
\begin{figure}
	\begin{subfigure}{0.48\textwidth}
		\scalebox{0.76}{
			\begin{tikzpicture}
	\begin{axis}[
		xlabel={$\overline{u} [m/s]$},
		ylabel={$Z [m]$},
		restrict x to domain = -0.2:0.2,
		restrict y to domain = 0.2:0.4,
		grid=none,
		yticklabel style={
			/pgf/number format/fixed,
			/pgf/number format/precision=5
		},
		scaled y ticks=false,
		scaled x ticks=false,
		xmin=-0.2,
		xmax=0.2,
		ymin= 0.2,
		ymax= 0.4,
		xtick = {-0.2,-0.1,0.0,0.1,0.2},
		ytick = {0.2,0.3,0.4},
		xticklabels = {-0.2,-0.1,0.0,0.1,0.2},
		ylabel style={rotate=-90},
		]
		
		\addplot [red,mark=o,mark size=4pt, mark repeat=15, line width=0.4mm] table [x=UMean:0, y=Points:2, col sep=comma] {images/spilling/tables/line_1.csv}; \label{velocity_red_line_spilling}
		
		\addplot [blue,mark=triangle*,mark size=4pt, mark repeat=1, only marks] table [x=u, y=z, col sep=comma] {images/spilling/tables/line_1_exp_vel.csv}; \label{velocity_blue_triangle_spilling}
		
		\addplot [black,mark=diamond,mark size=4pt, mark repeat=2, line width=0.4mm] table [x=u, y=z, col sep=comma] {images/spilling/tables/line_1_devolder_vel_kwsst.csv}; \label{velocity_black_line_spilling}
		
	\end{axis}
\end{tikzpicture}
		}
    \subcaption{$x=7.275$ m.}
	\end{subfigure}
	\begin{subfigure}{0.48\textwidth}
		\scalebox{0.75}{
			\begin{tikzpicture}
	\begin{axis}[
		xlabel={$\overline{u} [m/s]$},
		ylabel={$Z [m]$},
		restrict x to domain = -0.2:0.2,
		restrict y to domain = 0.2:0.4,
		grid=none,
		yticklabel style={
			/pgf/number format/fixed,
			/pgf/number format/precision=5
		},
		scaled y ticks=false,
		scaled x ticks=false,
		xmin=-0.2,
		xmax=0.2,
		ymin= 0.2,
		ymax= 0.4,
		xtick = {-0.2,-0.1,0.0,0.1,0.2},
		ytick = {0.2,0.3,0.4},
		xticklabels = {-0.2,-0.1,0.0,0.1,0.2},
		ylabel style={rotate=-90},
		]
		
		\addplot [red,mark=o,mark size=4pt, mark repeat=15, line width=0.4mm] table [x=UMean:0, y=Points:2, col sep=comma] {images/spilling/tables/line_2.csv};
		
		\addplot [blue,mark=triangle*,mark size=4pt, mark repeat=1, only marks] table [x=u, y=z, col sep=comma] {images/spilling/tables/line_2_exp_vel.csv};
		
		\addplot [black,mark=diamond,mark size=4pt, mark repeat=2, line width=0.4mm] table [x=u, y=z, col sep=comma] {images/spilling/tables/line_2_devolder_vel_kwsst.csv};
		
	\end{axis}
\end{tikzpicture}
		}
    \subcaption{$x=7.885$ m.}
	\end{subfigure}
	\hfill
	\begin{subfigure}{0.48\textwidth}
		\scalebox{0.76}{
			\begin{tikzpicture}
	\begin{axis}[
		xlabel={$\overline{u} [m/s]$},
		ylabel={$Z [m]$},
	restrict x to domain = -0.2:0.2,
		restrict y to domain = 0.2:0.4,
		grid=none,
		yticklabel style={
			/pgf/number format/fixed,
			/pgf/number format/precision=5
		},
		scaled y ticks=false,
		scaled x ticks=false,
		xmin=-0.2,
		xmax=0.2,
		ymin= 0.2,
		ymax= 0.4,
		xtick = {-0.2,-0.1,0.0,0.1,0.2},
		ytick = {0.2,0.3,0.4},
		xticklabels = {-0.2,-0.1,0.0,0.1,0.2},
		ylabel style={rotate=-90},
		]
		
		\addplot [red,mark=o,mark size=4pt, mark repeat=15, line width=0.4mm] table [x=UMean:0, y=Points:2, col sep=comma] {images/spilling/tables/line_3.csv};
		
		\addplot [blue,mark=triangle*,mark size=4pt, mark repeat=1, only marks] table [x=u, y=z, col sep=comma] {images/spilling/tables/line_3_exp_vel.csv};
		
		\addplot [black,mark=diamond,mark size=4pt, mark repeat=2, line width=0.4mm] table [x=u, y=z, col sep=comma] {images/spilling/tables/line_3_devolder_vel_kwsst.csv};
		
	\end{axis}
\end{tikzpicture}
		}
    \subcaption{$x=8.495$ m.}
	\end{subfigure}
	\begin{subfigure}{0.48\textwidth}
		\scalebox{0.75}{
			\begin{tikzpicture}
	\begin{axis}[
		xlabel={$\overline{u} [m/s]$},
		ylabel={$Z [m]$},
		restrict x to domain = -0.2:0.2,
		restrict y to domain = 0.2:0.4,
		grid=none,
		yticklabel style={
			/pgf/number format/fixed,
			/pgf/number format/precision=5
		},
		scaled y ticks=false,
		scaled x ticks=false,
		xmin=-0.2,
		xmax=0.2,
		ymin= 0.2,
		ymax= 0.4,
		xtick = {-0.2,-0.1,0.0,0.1,0.2},
		ytick = {0.2,0.3,0.4},
		xticklabels = {-0.2,-0.1,0.0,0.1,0.2},
		ylabel style={rotate=-90},
		]
		
		\addplot [red,mark=o,mark size=4pt, mark repeat=15, line width=0.4mm] table [x=UMean:0, y=Points:2, col sep=comma] {images/spilling/tables/line_4.csv};
		
		\addplot [blue,mark=triangle*,mark size=4pt, mark repeat=1, only marks] table [x=u, y=z, col sep=comma] {images/spilling/tables/line_4_exp_vel.csv};
		
		\addplot [black,mark=diamond,mark size=4pt, mark repeat=2, line width=0.4mm] table [x=u, y=z, col sep=comma] {images/spilling/tables/line_4_devolder_vel_kwsst.csv};
		
	\end{axis}
\end{tikzpicture}
		}
    \subcaption{$x=9.110$ m.}
	\end{subfigure}
	\hfill
	\begin{subfigure}{0.48\textwidth}
		\scalebox{0.75}{
			\begin{tikzpicture}
	\begin{axis}[
		xlabel={$\overline{u} [m/s]$},
		ylabel={$Z [m]$},
		restrict x to domain = -0.2:0.2,
		restrict y to domain = 0.2:0.4,
		grid=none,
		yticklabel style={
			/pgf/number format/fixed,
			/pgf/number format/precision=5
		},
		scaled y ticks=false,
		scaled x ticks=false,
		xmin=-0.2,
		xmax=0.2,
		ymin= 0.2,
		ymax= 0.4,
		xtick = {-0.2,-0.1,0.0,0.1,0.2},
		ytick = {0.2,0.3,0.4},
		xticklabels = {-0.2,-0.1,0.0,0.1,0.2},
		ylabel style={rotate=-90},
		]
		
		\addplot [red,mark=o,mark size=4pt, mark repeat=15, line width=0.4mm] table [x=UMean:0, y=Points:2, col sep=comma] {images/spilling/tables/line_5.csv};
		
		\addplot [blue,mark=triangle*,mark size=4pt, mark repeat=1, only marks] table [x=u, y=z, col sep=comma] {images/spilling/tables/line_5_exp_vel.csv};
		
		\addplot [black,mark=diamond,mark size=4pt, mark repeat=2, line width=0.4mm] table [x=u, y=z, col sep=comma] {images/spilling/tables/line_5_devolder_vel_kwsst.csv};
		
	\end{axis}
\end{tikzpicture}
		}
    \subcaption{$x=9.725$ m.}
	\end{subfigure}
	\caption{Mean horizontal velocity measured at five different positions for the spilling breakers. The \ref{velocity_red_line_spilling} is the presented model, \ref{velocity_blue_triangle_spilling} is experimental measurement from \cite{ting}, and \ref{velocity_black_line_spilling} is the numerical results by \cite{devolder} using $k\omega SST$ turbulence model.}
	\label{vel_spilling}
\end{figure}
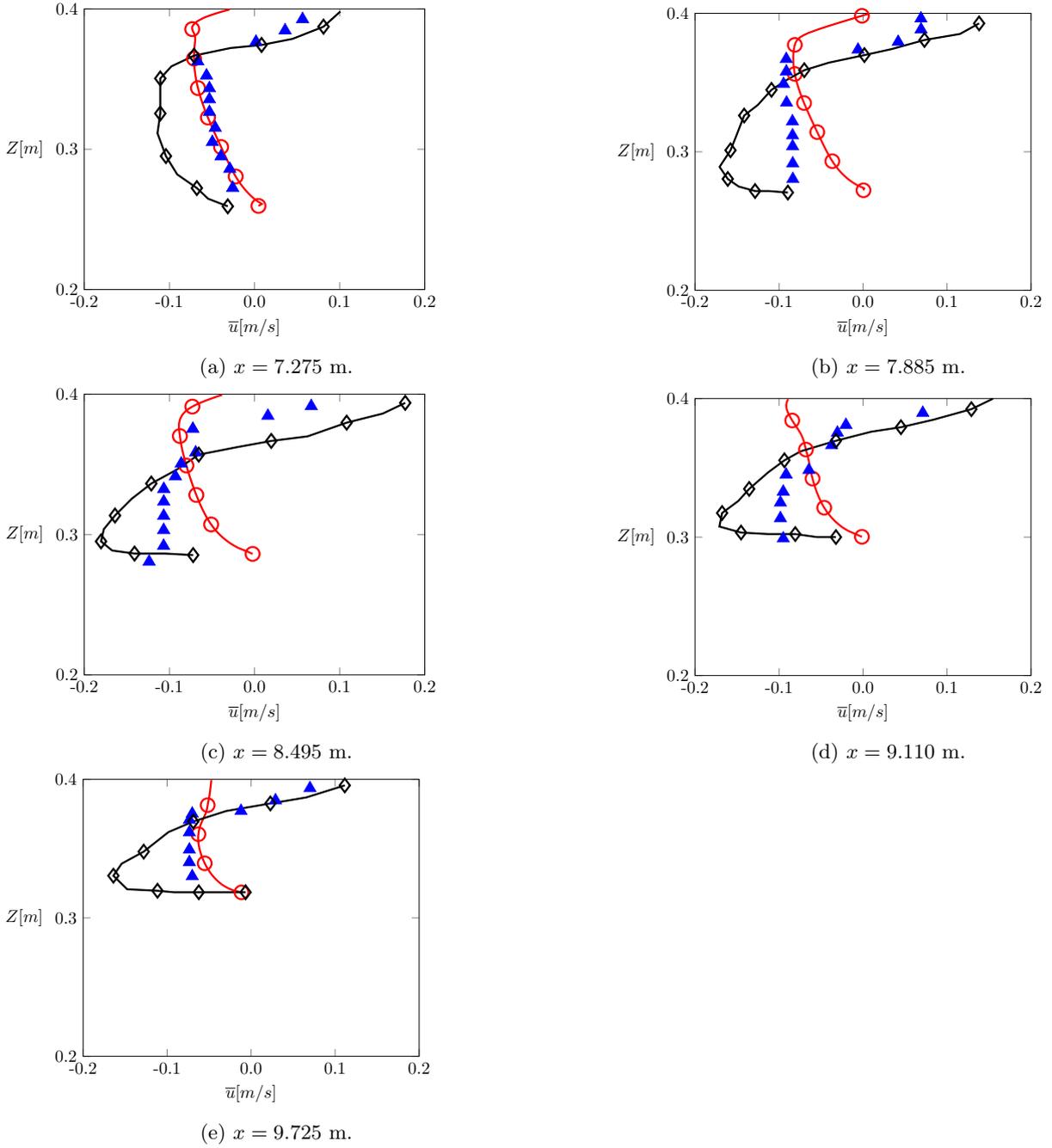
\begin{figure}
	\begin{subfigure}{0.48\textwidth}
		\scalebox{0.76}{
			\begin{tikzpicture}
	\begin{axis}[
		xlabel={$\overline{u} [m/s]$},
		ylabel={$Z [m]$},
		restrict x to domain = -0.2:0.2,
		restrict y to domain = 0.2:0.4,
		grid=none,
		yticklabel style={
			/pgf/number format/fixed,
			/pgf/number format/precision=5
		},
		scaled y ticks=false,
		scaled x ticks=false,
		xmin=-0.2,
		xmax=0.2,
		ymin= 0.2,
		ymax= 0.4,
		xtick = {-0.2,-0.1,0.0,0.1,0.2},
		ytick = {0.2,0.3,0.4},
		xticklabels = {-0.2,-0.1,0.0,0.1,0.2},
		ylabel style={rotate=-90},
		]
		
		\addplot [red,mark=o,mark size=4pt, mark repeat=15, line width=0.4mm] table [x=UMean:0, y=Points:2, col sep=comma] {images/plunging/tables/line_1.csv}; \label{velocity_red_line_plunging}
		
		\addplot [blue,mark=triangle*,mark size=4pt, mark repeat=1, only marks] table [x=u, y=z, col sep=comma] {images/plunging/tables/line_1_exp_vel.csv}; \label{velocity_blue_triangle_plunging}
		
		\addplot [black,mark=diamond,mark size=4pt, mark repeat=2, line width=0.4mm] table [x=u, y=z, col sep=comma] {images/plunging/tables/line_1_devolder_vel_kwsst.csv};
		\label{velocity_black_diamond_plunging}
		
	\end{axis}
\end{tikzpicture}
		}
    \subcaption{$x=7.275$ m.}
	\end{subfigure}
	\begin{subfigure}{0.48\textwidth}
		\scalebox{0.75}{
			\begin{tikzpicture}
	\begin{axis}[
		xlabel={$\overline{u} [m/s]$},
		ylabel={$Z [m]$},
		restrict x to domain = -0.2:0.2,
		restrict y to domain = 0.2:0.4,
		grid=none,
		yticklabel style={
			/pgf/number format/fixed,
			/pgf/number format/precision=5
		},
		scaled y ticks=false,
		scaled x ticks=false,
		xmin=-0.2,
		xmax=0.2,
		ymin= 0.2,
		ymax= 0.4,
		xtick = {-0.2,-0.1,0.0,0.1,0.2},
		ytick = {0.2,0.3,0.4},
		xticklabels = {-0.2,-0.1,0.0,0.1,0.2},
		ylabel style={rotate=-90},
		]
		
		\addplot [red,mark=o,mark size=4pt, mark repeat=15, line width=0.4mm] table [x=UMean:0, y=Points:2, col sep=comma] {images/plunging/tables/line_2.csv}; 
		
		\addplot [blue,mark=triangle*,mark size=4pt, mark repeat=1, only marks] table [x=u, y=z, col sep=comma] {images/plunging/tables/line_2_exp_vel.csv};
		
		\addplot [black,mark=diamond,mark size=4pt, mark repeat=2, line width=0.4mm] table [x=u, y=z, col sep=comma] {images/plunging/tables/line_2_devolder_vel_kwsst.csv};
		
	\end{axis}
\end{tikzpicture}
		}
    \subcaption{$x=7.885$ m.}
	\end{subfigure}
	\hfill
	\begin{subfigure}{0.48\textwidth}
		\scalebox{0.76}{
			\begin{tikzpicture}
	\begin{axis}[
		xlabel={$\overline{u} [m/s]$},
		ylabel={$Z [m]$},
		restrict x to domain = -0.2:0.2,
		restrict y to domain = 0.2:0.4,
		grid=none,
		yticklabel style={
			/pgf/number format/fixed,
			/pgf/number format/precision=5
		},
		scaled y ticks=false,
		scaled x ticks=false,
		xmin=-0.2,
		xmax=0.2,
		ymin= 0.2,
		ymax= 0.4,
		xtick = {-0.2,-0.1,0.0,0.1,0.2},
		ytick = {0.2,0.3,0.4},
		xticklabels = {-0.2,-0.1,0.0,0.1,0.2},
		ylabel style={rotate=-90},
		]
		
		\addplot [red,mark=o,mark size=4pt, mark repeat=15, line width=0.4mm] table [x=UMean:0, y=Points:2, col sep=comma] {images/plunging/tables/line_3.csv};
		
		\addplot [blue,mark=triangle*,mark size=4pt, mark repeat=1, only marks] table [x=u, y=z, col sep=comma] {images/spilling/tables/line_3_exp_vel.csv};
		
		\addplot [black,mark=diamond,mark size=4pt, mark repeat=2, line width=0.4mm] table [x=u, y=z, col sep=comma] {images/plunging/tables/line_3_devolder_vel_kwsst.csv};
		
	\end{axis}
\end{tikzpicture}
		}
    \subcaption{$x=8.495$ m.}
	\end{subfigure}
	\begin{subfigure}{0.48\textwidth}
		\scalebox{0.75}{
			\begin{tikzpicture}
	\begin{axis}[
		xlabel={$\overline{u} [m/s]$},
		ylabel={$Z [m]$},
		restrict x to domain = -0.2:0.2,
		restrict y to domain = 0.2:0.4,
		grid=none,
		yticklabel style={
			/pgf/number format/fixed,
			/pgf/number format/precision=5
		},
		scaled y ticks=false,
		scaled x ticks=false,
		xmin=-0.2,
		xmax=0.2,
		ymin= 0.2,
		ymax= 0.4,
		xtick = {-0.2,-0.1,0.0,0.1,0.2},
		ytick = {0.2,0.3,0.4},
		xticklabels = {-0.2,-0.1,0.0,0.1,0.2},
		ylabel style={rotate=-90},
		]
		
		\addplot [red,mark=o,mark size=4pt, mark repeat=15, line width=0.4mm] table [x=UMean:0, y=Points:2, col sep=comma] {images/plunging/tables/line_4.csv};
		
		\addplot [blue,mark=triangle*,mark size=4pt, mark repeat=1, only marks] table [x=u, y=z, col sep=comma] {images/spilling/tables/line_4_exp_vel.csv};
		
		\addplot [black,mark=diamond,mark size=4pt, mark repeat=2, line width=0.4mm] table [x=u, y=z, col sep=comma] {images/plunging/tables/line_4_devolder_vel_kwsst.csv};
		
	\end{axis}
\end{tikzpicture}
		}
    \subcaption{$x=9.110$ m.}
	\end{subfigure}
	\hfill
	\begin{subfigure}{0.48\textwidth}
		\scalebox{0.75}{
			\begin{tikzpicture}
	\begin{axis}[
		xlabel={$\overline{u} [m/s]$},
		ylabel={$Z [m]$},
		restrict x to domain = -0.2:0.2,
		restrict y to domain = 0.2:0.4,
		grid=none,
		yticklabel style={
			/pgf/number format/fixed,
			/pgf/number format/precision=5
		},
		scaled y ticks=false,
		scaled x ticks=false,
		xmin=-0.2,
		xmax=0.2,
		ymin= 0.2,
		ymax= 0.4,
		xtick = {-0.2,-0.1,0.0,0.1,0.2},
		ytick = {0.2,0.3,0.4},
		xticklabels = {-0.2,-0.1,0.0,0.1,0.2},
		ylabel style={rotate=-90},
		]
		
		\addplot [red,mark=o,mark size=4pt, mark repeat=15, line width=0.4mm] table [x=UMean:0, y=Points:2, col sep=comma] {images/plunging/tables/line_5.csv};
		
		\addplot [blue,mark=triangle*,mark size=4pt, mark repeat=1, only marks] table [x=u, y=z, col sep=comma] {images/plunging/tables/line_5_exp_vel.csv};
		
		\addplot [black,mark=diamond,mark size=4pt, mark repeat=2, line width=0.4mm] table [x=u, y=z, col sep=comma] {images/plunging/tables/line_5_devolder_vel_kwsst.csv};
		
	\end{axis}
\end{tikzpicture}
		}
    \subcaption{$x=9.725$ m.}
	\end{subfigure}
	\caption{Mean horizontal velocity measured at five different positions for the plunging breakers. The \ref{velocity_red_line_plunging} is the presented model, \ref{velocity_blue_triangle_plunging} is experimental measurement from \cite{ting}, and \ref{velocity_black_diamond_plunging}  is the numerical results by \cite{devolder} using $k\omega SST$ turbulence model.}
	\label{vel_plunging}
\end{figure}
Figures \ref{tke_spilling} and \ref{tke_plunging} show the mean turbulent kinetic energy on the horizontal axis and the height on the vertical axis. At the bottom (wall), it is found that the presented model predict higher TKE in comparison to the results of \cite{devolder}. The higher turbulence arises from: (1) the use of Neuman boundary condition of the TKE rather than a pre-specified value using wall functions and (2) The vorticity of the water moving in the negative direction after reaching the end of the domain. Furthermore, the presented model predicts a better agreement with the experimental measurement. The variation of the undertow TKE before the breaking point is minimum. The TKE is higher after the breaking point due to wave breaking. The presented model shows a steep variation of the TKE above the mean air-water interface which is compensated by a steep variation of the dissipation frequency (the jump in kinematic viscosity). Hence, the TKE does not grow rapidly below the air-water interface similar to the traditional $k \omega$ model.
\newline
\begin{figure}
	\begin{subfigure}{0.48\textwidth}
		\scalebox{0.76}{
			\begin{tikzpicture}
	\begin{axis}[
		xlabel={$\overline{k} [m^2/s^2]$},
		ylabel={$Z [m]$},
		restrict x to domain = 0:0.02,
		restrict y to domain = 0.2:0.4,
		grid=none,
		yticklabel style={
			/pgf/number format/fixed,
			/pgf/number format/precision=5
		},
		scaled y ticks=false,
		scaled x ticks=false,
		xmin=0,
		xmax=0.02,
		ymin= 0.2,
		ymax= 0.4,
		xtick = {0,0.005,0.01,0.015,0.02},
		ytick = {0.2,0.3,0.4},
		xticklabels = {0,0.005,0.01,0.015,0.02},
		ylabel style={rotate=-90},
		]
		
		\addplot [red,mark=o,mark size=4pt, mark repeat=15, line width=0.4mm] table [x=kMean, y=Points:2, col sep=comma] {images/spilling/tables/line_1.csv}; \label{red_line_spilling}

		\addplot [blue,mark=triangle*,mark size=4pt, mark repeat=2, only marks] table [x=k, y=z, col sep=comma] {images/spilling/tables/line_1_exp.csv}; \label{blue_triangle_spilling}
		
		\addplot [black,mark=diamond,mark size=4pt, mark repeat=2, line width=0.4mm] table [x=k, y=z, col sep=comma] {images/spilling/tables/line_1_devolder_tke_kwsst.csv}; \label{black_line_spilling}

	\end{axis}
\end{tikzpicture}
		}
    \subcaption{$x=8.345$ m.}
	\end{subfigure}
	\begin{subfigure}{0.48\textwidth}
		\scalebox{0.75}{
			\begin{tikzpicture}
	\begin{axis}[
		xlabel={$\overline{k} [m^2/s^2]$},
		ylabel={$Z [m]$},
		restrict x to domain = 0:0.02,
		restrict y to domain = 0.2:0.4,
		grid=none,
		yticklabel style={
			/pgf/number format/fixed,
			/pgf/number format/precision=5
		},
		scaled y ticks=false,
		scaled x ticks=false,
		xmin=0,
		xmax=0.02,
		ymin= 0.2,
		ymax= 0.4,
		xtick = {0,0.005,0.01,0.015,0.02},
		ytick = {0.2,0.3,0.4},
		xticklabels = {0,0.005,0.01,0.015,0.02},
		ylabel style={rotate=-90},
		]
		
		\addplot [red,mark=o,mark size=4pt, mark repeat=15, line width=0.4mm] table [x=kMean, y=Points:2, col sep=comma] {images/spilling/tables/line_2.csv};
		
		\addplot [blue,mark=triangle*,mark size=4pt, mark repeat=2, only marks] table [x=k, y=z, col sep=comma] {images/spilling/tables/line_2_exp.csv};
		
		\addplot [black,mark=diamond,mark size=4pt, mark repeat=2, line width=0.4mm] table [x=k, y=z, col sep=comma] {images/spilling/tables/line_2_devolder_tke_kwsst.csv};
		
	\end{axis}
\end{tikzpicture}
		}
    \subcaption{$x=8.795$ m.}
	\end{subfigure}
	\hfill
	\begin{subfigure}{0.48\textwidth}
		\scalebox{0.76}{
			\begin{tikzpicture}
	\begin{axis}[
		xlabel={$\overline{k} [m^2/s^2]$},
		ylabel={$Z [m]$},
		restrict x to domain = 0:0.02,
		restrict y to domain = 0.2:0.4,
		grid=none,
		yticklabel style={
			/pgf/number format/fixed,
			/pgf/number format/precision=5
		},
		scaled y ticks=false,
		scaled x ticks=false,
		xmin=0,
		xmax=0.02,
		ymin= 0.2,
		ymax= 0.4,
		xtick = {0,0.005,0.01,0.015,0.02},
		ytick = {0.2,0.3,0.4},
		xticklabels = {0,0.005,0.01,0.015,0.02},
		ylabel style={rotate=-90},
		]
		
		\addplot [red,mark=o,mark size=4pt, mark repeat=15, line width=0.4mm] table [x=kMean, y=Points:2, col sep=comma] {images/spilling/tables/line_3.csv};
		
		\addplot [blue,mark=triangle*,mark size=4pt, mark repeat=2, only marks] table [x=k, y=z, col sep=comma] {images/spilling/tables/line_3_exp.csv};
		
		\addplot [black,mark=diamond,mark size=4pt, mark repeat=2, line width=0.4mm] table [x=k, y=z, col sep=comma] {images/spilling/tables/line_3_devolder_tke_kwsst.csv};
		
	\end{axis}
\end{tikzpicture}
		}
    \subcaption{$x=9.295$ m.}
	\end{subfigure}
	\begin{subfigure}{0.48\textwidth}
		\scalebox{0.75}{
			\begin{tikzpicture}
	\begin{axis}[
		xlabel={$\overline{k} [m^2/s^2]$},
		ylabel={$Z [m]$},
		restrict x to domain = 0:0.02,
		restrict y to domain = 0.2:0.4,
		grid=none,
		yticklabel style={
			/pgf/number format/fixed,
			/pgf/number format/precision=5
		},
		scaled y ticks=false,
		scaled x ticks=false,
		xmin=0,
		xmax=0.02,
		ymin= 0.2,
		ymax= 0.4,
		xtick = {0,0.005,0.01,0.015,0.02},
		ytick = {0.2,0.3,0.4},
		xticklabels = {0,0.005,0.01,0.015,0.02},
		ylabel style={rotate=-90},
		]
		
		\addplot [red,mark=o,mark size=4pt, mark repeat=15, line width=0.4mm] table [x=kMean, y=Points:2, col sep=comma] {images/spilling/tables/line_4.csv};
		
		\addplot [blue,mark=triangle*,mark size=4pt, mark repeat=2, only marks] table [x=k, y=z, col sep=comma] {images/spilling/tables/line_4_exp.csv};
		
		\addplot [black,mark=diamond,mark size=4pt, mark repeat=2, line width=0.4mm] table [x=k, y=z, col sep=comma] {images/spilling/tables/line_4_devolder_tke_kwsst.csv};
		
	\end{axis}
\end{tikzpicture}
		}
    \subcaption{$x=9.795$ m.}
	\end{subfigure}
	\hfill
	\begin{subfigure}{0.48\textwidth}
		\scalebox{0.75}{
			\begin{tikzpicture}
	\begin{axis}[
		xlabel={$\overline{k} [m^2/s^2]$},
		ylabel={$Z [m]$},
		restrict x to domain = 0:0.02,
		restrict y to domain = 0.2:0.4,
		grid=none,
		yticklabel style={
			/pgf/number format/fixed,
			/pgf/number format/precision=5
		},
		scaled y ticks=false,
		scaled x ticks=false,
		xmin=0,
		xmax=0.02,
		ymin= 0.2,
		ymax= 0.4,
		xtick = {0,0.005,0.01,0.015,0.02},
		ytick = {0.2,0.3,0.4},
		xticklabels = {0,0.005,0.01,0.015,0.02},
		ylabel style={rotate=-90},
		]
		
		\addplot [red,mark=o,mark size=4pt, mark repeat=15, line width=0.4mm] table [x=kMean, y=Points:2, col sep=comma] {images/spilling/tables/line_5.csv};
		
		\addplot [blue,mark=triangle*,mark size=4pt, mark repeat=2, only marks] table [x=k, y=z, col sep=comma] {images/spilling/tables/line_5_exp.csv};
		
		\addplot [black,mark=diamond,mark size=4pt, mark repeat=2, line width=0.4mm] table [x=k, y=z, col sep=comma] {images/spilling/tables/line_5_devolder_tke_kwsst.csv};
		
	\end{axis}
\end{tikzpicture}
		}
    \subcaption{$x=10.395$ m.}
	\end{subfigure}
	\caption{Mean turbulent kinetic energy measured at five different positions for the spilling breakers. The \ref{red_line_spilling} is the presented model, \ref{blue_triangle_spilling} is experimental measurement from \cite{ting}, and \ref{black_line_spilling} is the numerical results by \cite{devolder} using $k\omega SST$ turbulence model.}
	\label{tke_spilling}
\end{figure}
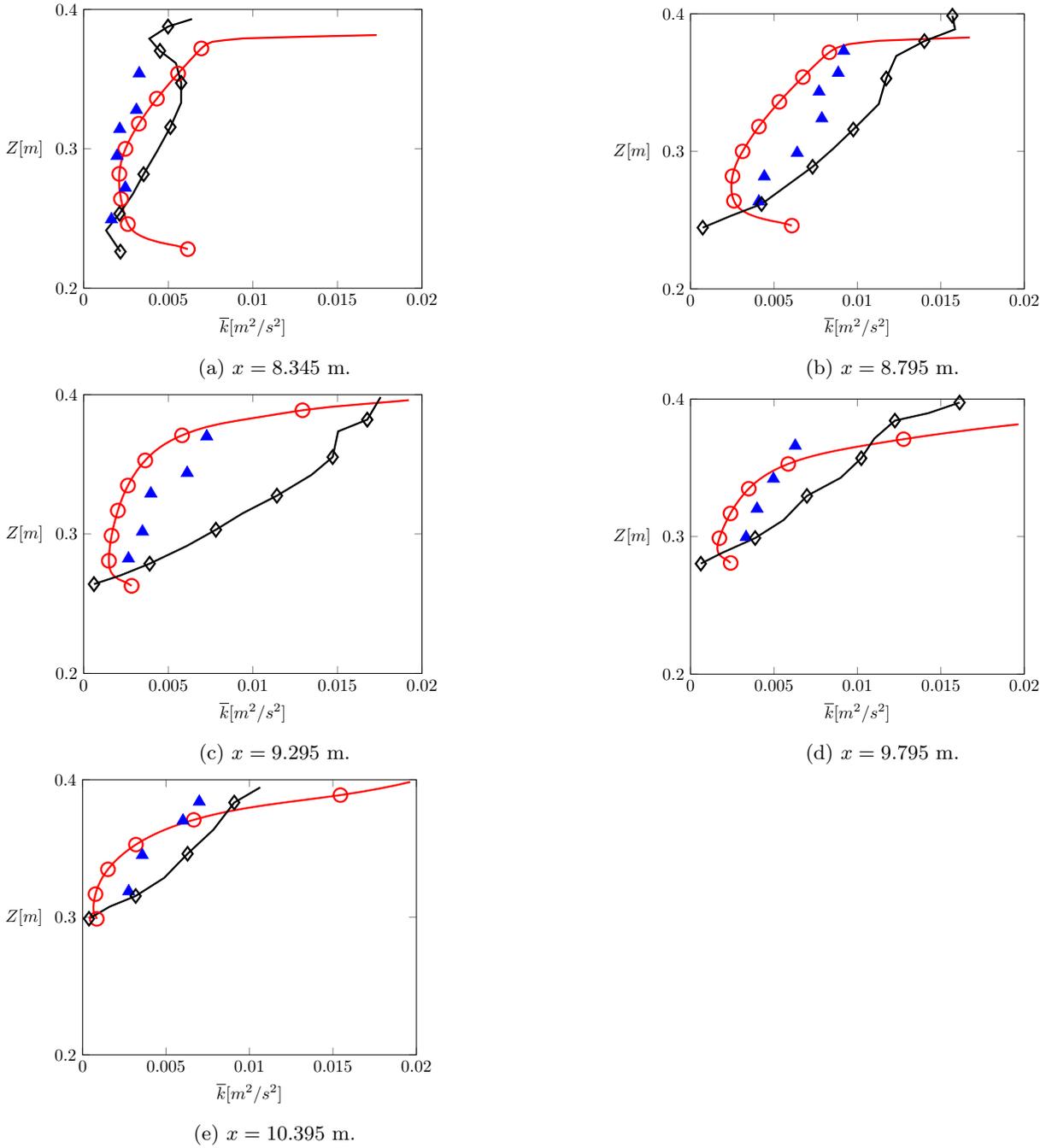

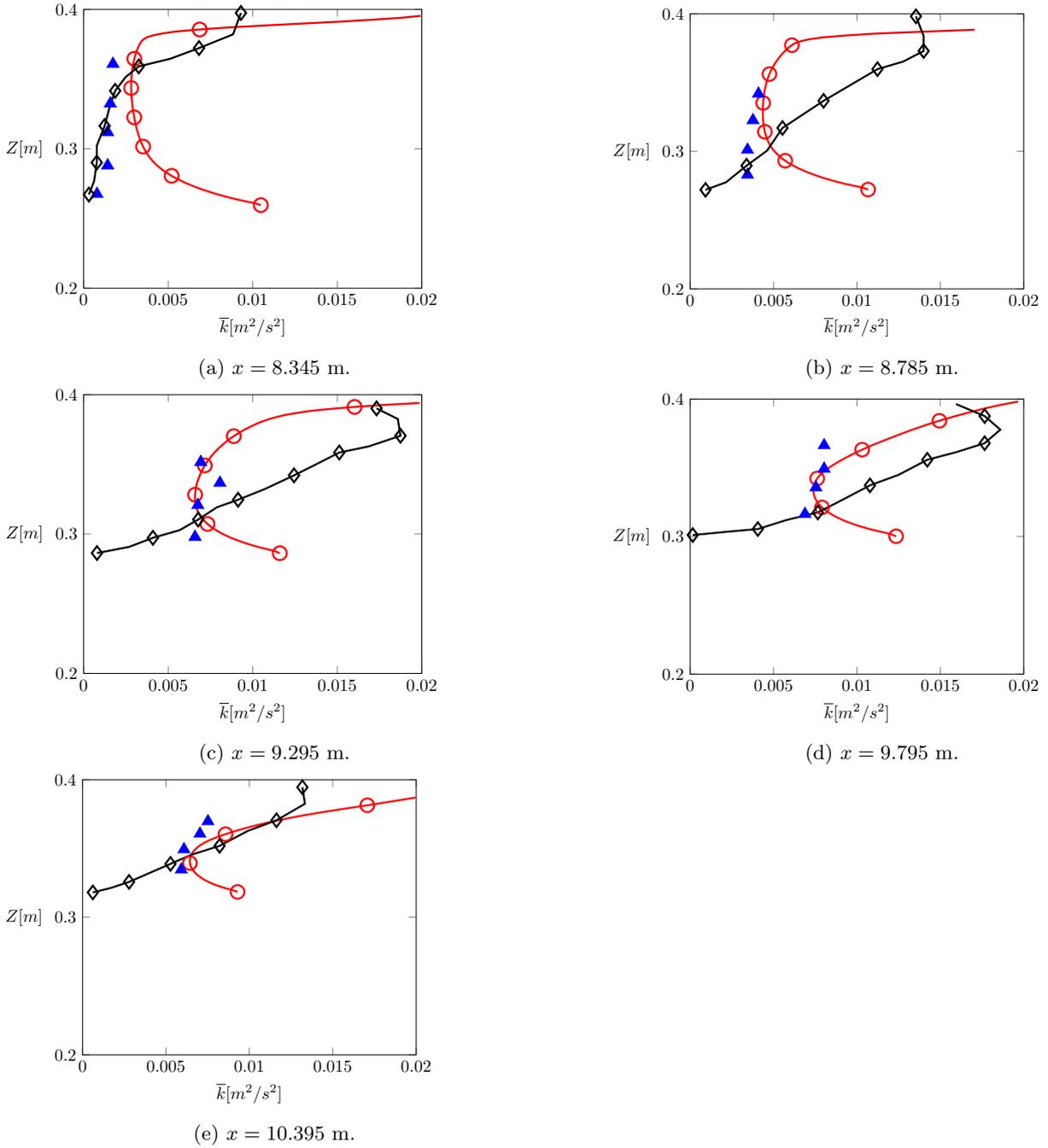
\begin{figure}
	\begin{subfigure}{0.48\textwidth}
		\scalebox{0.76}{
			\begin{tikzpicture}
	\begin{axis}[
		xlabel={$\overline{k} [m^2/s^2]$},
		ylabel={$Z [m]$},
		restrict x to domain = 0:0.02,
		restrict y to domain = 0.2:0.4,
		grid=none,
		yticklabel style={
			/pgf/number format/fixed,
			/pgf/number format/precision=5
		},
		scaled y ticks=false,
		scaled x ticks=false,
		xmin=0,
		xmax=0.02,
		ymin= 0.2,
		ymax= 0.4,
		xtick = {0,0.005,0.01,0.015,0.02},
		ytick = {0.2,0.3,0.4},
		xticklabels = {0,0.005,0.01,0.015,0.02},
		ylabel style={rotate=-90},
		]
		
		\addplot [red,mark=o,mark size=4pt, mark repeat=15, line width=0.4mm] table [x=kMean, y=Points:2, col sep=comma] {images/plunging/tables/line_1.csv}; \label{red_line_plunging}
		
		\addplot [blue,mark=triangle*,mark size=4pt, mark repeat=2, only marks] table [x=k, y=z, col sep=comma] {images/plunging/tables/line_1_exp.csv}; \label{blue_triangle_plunging}
		
		\addplot [black,mark=diamond,mark size=4pt, mark repeat=2, line width=0.4mm] table [x=k, y=z, col sep=comma] {images/plunging/tables/line_1_devolder_tke_kwsst.csv}; \label{black_line_plunging}
		
	\end{axis}
\end{tikzpicture}
		}
    \subcaption{$x=8.345$ m.}
	\end{subfigure}
	\begin{subfigure}{0.48\textwidth}
		\scalebox{0.75}{
			\begin{tikzpicture}
	\begin{axis}[
		xlabel={$\overline{k} [m^2/s^2]$},
		ylabel={$Z [m]$},
		restrict x to domain = 0:0.02,
		restrict y to domain = 0.2:0.4,
		grid=none,
		yticklabel style={
			/pgf/number format/fixed,
			/pgf/number format/precision=5
		},
		scaled y ticks=false,
		scaled x ticks=false,
		xmin=0,
		xmax=0.02,
		ymin= 0.2,
		ymax= 0.4,
		xtick = {0,0.005,0.01,0.015,0.02},
		ytick = {0.2,0.3,0.4},
		xticklabels = {0,0.005,0.01,0.015,0.02},
		ylabel style={rotate=-90},
		]
		
		\addplot [red,mark=o,mark size=4pt, mark repeat=15, line width=0.4mm] table [x=kMean, y=Points:2, col sep=comma] {images/plunging/tables/line_2.csv};
		
		\addplot [blue,mark=triangle*,mark size=4pt, mark repeat=2, only marks] table [x=k, y=z, col sep=comma] {images/plunging/tables/line_2_exp.csv};
		
		\addplot [black,mark=diamond,mark size=4pt, mark repeat=2, line width=0.4mm] table [x=k, y=z, col sep=comma] {images/plunging/tables/line_2_devolder_tke_kwsst.csv};
		
	\end{axis}
\end{tikzpicture}
		}
    \subcaption{$x=8.785$ m.}
	\end{subfigure}
	\hfill
	\begin{subfigure}{0.48\textwidth}
		\scalebox{0.76}{
			\begin{tikzpicture}
	\begin{axis}[
		xlabel={$\overline{k} [m^2/s^2]$},
		ylabel={$Z [m]$},
		restrict x to domain = 0:0.02,
		restrict y to domain = 0.2:0.4,
		grid=none,
		yticklabel style={
			/pgf/number format/fixed,
			/pgf/number format/precision=5
		},
		scaled y ticks=false,
		scaled x ticks=false,
		xmin=0,
		xmax=0.02,
		ymin= 0.2,
		ymax= 0.4,
		xtick = {0,0.005,0.01,0.015,0.02},
		ytick = {0.2,0.3,0.4},
		xticklabels = {0,0.005,0.01,0.015,0.02},
		ylabel style={rotate=-90},
		]
		
		\addplot [red,mark=o,mark size=4pt, mark repeat=15, line width=0.4mm] table [x=kMean, y=Points:2, col sep=comma] {images/plunging/tables/line_3.csv};
		
		\addplot [blue,mark=triangle*,mark size=4pt, mark repeat=2, only marks] table [x=k, y=z, col sep=comma] {images/plunging/tables/line_3_exp.csv};
		
		\addplot [black,mark=diamond,mark size=4pt, mark repeat=2, line width=0.4mm] table [x=k, y=z, col sep=comma] {images/plunging/tables/line_3_devolder_tke_kwsst.csv};
		
	\end{axis}
\end{tikzpicture}
		}
    \subcaption{$x=9.295$ m.}
	\end{subfigure}
	\begin{subfigure}{0.48\textwidth}
		\scalebox{0.75}{
			\begin{tikzpicture}
	\begin{axis}[
		xlabel={$\overline{k} [m^2/s^2]$},
		ylabel={$Z [m]$},
		restrict x to domain = 0:0.02,
		restrict y to domain = 0.2:0.4,
		grid=none,
		yticklabel style={
			/pgf/number format/fixed,
			/pgf/number format/precision=5
		},
		scaled y ticks=false,
		scaled x ticks=false,
		xmin=0,
		xmax=0.02,
		ymin= 0.2,
		ymax= 0.4,
		xtick = {0,0.005,0.01,0.015,0.02},
		ytick = {0.2,0.3,0.4},
		xticklabels = {0,0.005,0.01,0.015,0.02},
		ylabel style={rotate=-90},
		]
		
		\addplot [red,mark=o,mark size=4pt, mark repeat=15, line width=0.4mm] table [x=kMean, y=Points:2, col sep=comma] {images/plunging/tables/line_4.csv};
		
		\addplot [blue,mark=triangle*,mark size=4pt, mark repeat=2, only marks] table [x=k, y=z, col sep=comma] {images/plunging/tables/line_4_exp.csv};
		
		\addplot [black,mark=diamond,mark size=4pt, mark repeat=2, line width=0.4mm] table [x=k, y=z, col sep=comma] {images/plunging/tables/line_4_devolder_tke_kwsst.csv};
		
	\end{axis}
\end{tikzpicture}
		}
    \subcaption{$x=9.795$ m.}
	\end{subfigure}
	\hfill
	\begin{subfigure}{0.48\textwidth}
		\scalebox{0.75}{
			\begin{tikzpicture}
	\begin{axis}[
		xlabel={$\overline{k} [m^2/s^2]$},
		ylabel={$Z [m]$},
		restrict x to domain = 0:0.02,
		restrict y to domain = 0.2:0.4,
		grid=none,
		yticklabel style={
			/pgf/number format/fixed,
			/pgf/number format/precision=5
		},
		scaled y ticks=false,
		scaled x ticks=false,
		xmin=0,
		xmax=0.02,
		ymin= 0.2,
		ymax= 0.4,
		xtick = {0,0.005,0.01,0.015,0.02},
		ytick = {0.2,0.3,0.4},
		xticklabels = {0,0.005,0.01,0.015,0.02},
		ylabel style={rotate=-90},
		]
		
		\addplot [red,mark=o,mark size=4pt, mark repeat=15, line width=0.4mm] table [x=kMean, y=Points:2, col sep=comma] {images/plunging/tables/line_5.csv};
		
		\addplot [blue,mark=triangle*,mark size=4pt, mark repeat=2, only marks] table [x=k, y=z, col sep=comma] {images/plunging/tables/line_5_exp.csv};
		
		\addplot [black,mark=diamond,mark size=4pt, mark repeat=2, line width=0.4mm] table [x=k, y=z, col sep=comma] {images/plunging/tables/line_5_devolder_tke_kwsst.csv};
		
	\end{axis}
\end{tikzpicture}
		}
    \subcaption{$x=10.395$ m.}
	\end{subfigure}
	\caption{Mean turbulent kinetic energy measured at five different positions for the plunging breakers. The \ref{red_line_plunging} is the presented model, \ref{blue_triangle_plunging} is experimental measurement from \cite{ting}, and \ref{black_line_plunging} is the numerical results by \cite{devolder} using $k\omega SST$ turbulence model.}
	\label{tke_plunging}
\end{figure}

\section{Conclusions}
\label{sec:conclusions}
In this paper, the family of two equations turbulence models (both $k\epsilon$ and $k\omega$) are extended for two-phase flows. First, the jump conditions are derived. From physical analysis, the ratio between the kinematic viscosity of the water and air is one order of magnitude. Thus, it is reasonable to assume that the turbulence time scale has the same ratio between the two phases i.e, turbulence is dissipated in air at a higher rate than water. The effect of time scale is considered assuming negligible effects of other jump conditions. Hence, a new field, inverse turbulence area $\zeta$ with units $\text{m}^{-2}$, appears naturally by dividing the turbulence frequency by the kinematic viscosity. The new model is formulated in terms of $k \zeta SST$ equipped with the same limiter functions of the $k\omega SST$ model. It is important to stress that the buoyancy modification term presented by \cite{devolder} is not utilized in this work. The model is tested at two wave conditions and the results are compared with experimental measurements. The proposed model predicted the undertow TKE accurately and is consistent with the measurements. It is also explains the over-production of TKE in previous research. The air above the air-water interface behaves a sink of the TKE. This leads to lower turbulence levels in water.

\bibliographystyle{plain}
\bibliography{references}

\begin{thebibliography}{10}

\bibitem{brown_evaluation_turb_surf_zone}
S.A. Brown, D.M. Greaves, V.~Magar, and D.C. Conley.
\newblock Evaluation of turbulence closure models under spilling and plunging
  breakers in the surf zone.
\newblock {\em Coastal Engineering}, 114:177--193, 2016.

\bibitem{foamStar1}
Youngmyung Choi, Benjamin Bouscasse, Sopheak Seng, Guillaume Ducrozet, Lionel
  Gentaz, and Pierre Ferrant.
\newblock Generation of regular and irregular waves in navier-stokes cfd
  solvers by matching with the nonlinear potential wave solution at the
  boundaries.
\newblock volume Volume 2: CFD and FSI of {\em International Conference on
  Offshore Mechanics and Arctic Engineering}, 06 2018.

\bibitem{devolder}
Brecht Devolder, Peter Troch, and Pieter Rauwoens.
\newblock Performance of a buoyancy-modified k-$\omega$ and k-$\omega$ sst
  turbulence model for simulating wave breaking under regular waves using
  openfoam.
\newblock {\em Coastal Engineering}, 138:49--65, 2018.

\bibitem{HIRT1981201}
C.W Hirt and B.D Nichols.
\newblock Volume of fluid (vof) method for the dynamics of free boundaries.
\newblock {\em Journal of Computational Physics}, 39(1):201--225, 1981.

\bibitem{waves2foam}
Niels~G. Jacobsen, David~R. Fuhrman, and Jørgen Fredsøe.
\newblock A wave generation toolbox for the open-source cfd library:
  Openfoam®.
\newblock {\em International Journal for Numerical Methods in Fluids},
  70(9):1073--1088, 2012.

\bibitem{jasakPhD}
H~Jasak.
\newblock {\em Error Analysis and Estimation for the Finite Volume Method with
  Applications to Fluid Flows}.
\newblock PhD thesis, Imperial college london, 1996.

\bibitem{rng}
W.P Jones and B.E Launder.
\newblock The prediction of laminarization with a two-equation model of
  turbulence.
\newblock {\em International Journal of Heat and Mass Transfer},
  15(2):301--314, 1972.

\bibitem{larsen_fuhrman_2018}
Bjarke~Eltard Larsen and David~R. Fuhrman.
\newblock On the over-production of turbulence beneath surface waves in
  reynolds-averaged navier–stokes models.
\newblock {\em Journal of Fluid Mechanics}, 853:419–460, 2018.

\bibitem{launder}
B.E. Launder and B.I. Sharma.
\newblock Application of the energy-dissipation model of turbulence to the
  calculation of flow near a spinning disc.
\newblock {\em Letters in Heat and Mass Transfer}, 1(2):131--137, 1974.

\bibitem{lin_liu_1998}
Pengzhi Lin and Philip L.-F. Liu.
\newblock A numerical study of breaking waves in the surf zone.
\newblock {\em Journal of Fluid Mechanics}, 359:239–264, 1998.

\bibitem{mayer}
Stefan Mayer and Per~A. Madsen.
\newblock {\em Simulation of Breaking Waves in the Surf Zone using a
  Navier-Stokes Solver}, pages 928--941.
\newblock 2000.

\bibitem{menter_1}
F.~Menter.
\newblock {\em Zonal Two Equation k-w Turbulence Models For Aerodynamic Flows}.
\newblock 1993.

\bibitem{menter_2}
F.~R. Menter.
\newblock Influence of freestream values on k-omega turbulence model
  predictions.
\newblock {\em AIAA Journal}, 30(6):1657--1659, 1992.

\bibitem{menter_3}
Florian Menter, M.~Kuntz, and RB~Langtry.
\newblock Ten years of industrial experience with the sst turbulence model.
\newblock {\em Heat and Mass Transfer}, 4, 01 2003.

\bibitem{pope}
Stephen~B. Pope.
\newblock {\em Turbulent Flows}.
\newblock Cambridge University Press, 2000.

\bibitem{prandtl}
L.~Prandtl.
\newblock 7. bericht über untersuchungen zur ausgebildeten turbulenz.
\newblock {\em ZAMM - Journal of Applied Mathematics and Mechanics /
  Zeitschrift für Angewandte Mathematik und Mechanik}, 5(2):136--139, 1925.

\bibitem{spalart}
P.~Spalart and S.~Allmaras.
\newblock {\em A one-equation turbulence model for aerodynamic flows}.
\newblock 1992.

\bibitem{hsu}
Hsu Tai-Wen, Hsieh Chih-Min andTsai Chin-Yen, and Ou~Shan-Hwei.
\newblock Coupling vof/plic and embedding method for simulating wave breaking
  on a sloping beach.
\newblock {\em Journal of Marine Science and Technology}, 23, 2015.

\bibitem{tennekes}
Henk Tennekes and John~L. Lumley.
\newblock {\em {A First Course in Turbulence}}.
\newblock The MIT Press, 03 1972.

\bibitem{ting}
Francis~C.K. Ting and James~T. Kirby.
\newblock Observation of undertow and turbulence in a laboratory surf zone.
\newblock {\em Coastal Engineering}, 24(1):51--80, 1994.

\bibitem{vanmale}
Karim {Van Maele} and Bart Merci.
\newblock Application of two buoyancy-modified k–$\epsilon$ turbulence models
  to different types of buoyant plumes.
\newblock {\em Fire Safety Journal}, 41(2):122--138, 2006.

\bibitem{of_paper}
H.~G. Weller, G.~Tabor, H.~Jasak, and C.~Fureby.
\newblock A tensorial approach to computational continuum mechanics using
  object-oriented techniques.
\newblock {\em Computers in Physics}, 12(6):620--631, 1998.

\bibitem{wilcox}
D.C. Wilcox.
\newblock {\em Turbulence Modeling for CFD}.
\newblock DCW Industries, Incorporated, 1994.

\bibitem{yakhot}
V.~Yakhot, S.~A. Orszag, S.~Thangam, T.~B. Gatski, and C.~G. Speziale.
\newblock Development of turbulence models for shear flows by a double
  expansion technique.
\newblock {\em Physics of Fluids A: Fluid Dynamics}, 4(7):1510--1520, 1992.

\bibitem{zalesakFCT}
Steven~T Zalesak.
\newblock Fully multidimensional flux-corrected transport algorithms for
  fluids.
\newblock {\em Journal of Computational Physics}, 31(3):335 -- 362, 1979.

\end{thebibliography}

\end{document}